\documentclass[aps,prd,twocolumn,superscriptaddress,amsmath,amssymb,nofootinbib,preprintnumbers,floatfix,longbibliography]{revtex4-1}
\usepackage[dvipsnames]{xcolor}
\usepackage[colorlinks]{hyperref}
\usepackage{xspace}
\usepackage{mathtools}
\usepackage{bm}
\usepackage{braket}
\usepackage{slashed}

\usepackage{tikz}
\usetikzlibrary{tikzmark}
\usetikzlibrary{positioning}
\usepackage[normalem]{ulem} 

\hypersetup{linkcolor=BrickRed,citecolor=Green,
filecolor=Mulberry,
urlcolor=NavyBlue,
menucolor=BrickRed,
runcolor=Mulberry
}

\definecolor{lime}{HTML}{A6CE39}
\DeclareRobustCommand{\orcidicon}{\hspace{-1mm}
	\begin{tikzpicture}
	\draw[lime, fill=lime] (0,0) 
	circle [radius=0.16] 
	node[white] {{\fontfamily{qag}\selectfont \tiny \,ID}};
	\draw[white, fill=white] (-0.0525,0.095) 
	circle [radius=0.007];
	\end{tikzpicture}
	\hspace{-3mm}
}

\foreach \x in {A, ..., Z}{\expandafter\xdef\csname orcid\x\endcsname{\noexpand\href{https://orcid.org/\csname orcidauthor\x\endcsname}
			{\noexpand\orcidicon}}
}

\usepackage{graphicx}
\usepackage{bbm}

\makeatletter
\newcommand{\abs}{\@ifstar\abssmall\absbig}
\newcommand{\absbig}[1]{\left \lvert #1 \right \rvert}
\newcommand{\abssmall}[1]{\lvert #1 \rvert}
\makeatother

\newcommand{\Tr}{\mathrm{Tr}}
\renewcommand{\Im}{\mathrm{Im}}
\renewcommand{\Re}{\mathrm{Re}}
\newcommand{\dd}{\mathrm{d}}

\newcommand{\vpp}{\vec{p}^{\prime}}

\renewcommand{\vec}{\mathbf}

\begin{document}
\preprint{N3AS-26-011}

\title{Neutrino helicity oscillations in astrophysical environments: a many-body approach}

\author{Yiheng Xu}
\email{y7xu@physics.ucsd.edu}
\affiliation{Department of Physics, University of California San Diego, La Jolla, CA 92093, USA}

\author{Julien Froustey\orcidJ{}}
\email{julien.froustey@ific.uv.es}
\affiliation{Institut de Física Corpuscular (IFIC), CSIC-Universitat de València, Parc Científic UV, C/ Catedrático José Beltrán 2, E-46980 Paterna (Valencia), Spain}

\author{George M.~Fuller\orcidG{}}
\email{gfuller@physics.ucsd.edu}
\affiliation{Department of Physics, University of California San Diego, La Jolla, CA 92093, USA}

\author{{Luk\'{a}\v{s} Gr\'{a}f}\orcidL{}}
\email{lukas.graf@matfyz.cuni.cz}
\affiliation{Institute of Particle and Nuclear Physics, Faculty of Mathematics and Physics, Charles University in Prague, V Holešovičkách 2, 180 00 Praha 8, Czech Republic}
\affiliation{Institute of Physics, Silesian University in Opava, Bezru{\v{c}}ovo n{\'a}m{\v{e}}st{\'i} 1150/13, 746 01 Opava, Czech Republic}

\author{Amol V.~Patwardhan\orcidA{}}
\email{apatwardhan@reed.edu}
\affiliation{Department of Physics, Reed College, Portland, OR 97202, USA}

\begin{abstract}
    Neutrino rest mass enables left-handed states to \lq\lq flip\rq\rq\ to right-handed states and {\it vice versa}. In-medium effects can enhance the probability for such spin-flip. We demonstrate that a full many-body calculation of this process in neutrino-dense environments can lead to spin-flip probabilities that exceed by orders of magnitude those calculated with mean-field treatments. We study simple configurations with a few neutrinos in well-defined momentum states, for which we show that the helicity conversion enhancement is connected to many-body momentum exchange. Such an effect would therefore be missed in a calculation that considers only forward processes. We speculate on the potential astrophysical implications of these results and the range of applicability of our calculation and its limitations.  
\end{abstract}

\maketitle

\section{Introduction}

Neutrinos possess rest masses that are tiny compared to those of the charged leptons and quarks in their respective elementary particle families. Nevertheless, these minuscule masses can have profound implications for the evolution of the cosmos. In this work, we study one intriguing consequence of neutrino rest mass: the possibility of in-medium inter-conversion of neutrino left-handed and right-handed states in astrophysical settings. 

We preface our discussion by pointing out that neutrinos play a central role in the dynamics, nucleosynthesis, and observable signals of dense astrophysical environments such as core-collapse supernovae and binary neutron star mergers~\cite{Janka:2012wk,Mezzacappa:2020oyq,Burrows:2020qrp,Radice:2020ddv,Foucart:2022bth}. For example, $99 \, \%$ of the gravitational binding energy released in the collapse of the core of a massive star to a proto-neutron star, some $10 \, \%$ of the \emph{rest mass} of the core, is pumped into seas of neutrinos. On a time scale of seconds, $\sim{10}^{58}$ neutrinos, with an average energy $\sim{10}\, \mathrm{MeV}$, will diffuse out of the proto-neutron star and stream through the envelope of the star. Charged-current neutrino-nucleon processes, principally 
\begin{equation}
    \nu_e+n \rightleftharpoons p+e^- \ \ \text{and} \ \ 
    \bar\nu_e+p \rightleftharpoons n+e^+ \, ,
\label{eq:CCnp}    
\end{equation}
can be significant in transferring energy to the nascent core bounce-generated shock wave and in setting the neutron-to-proton ratio in any neutrino-heated outflow. That, in turn, has implications for nucleosynthesis pathways such as the $r$-process, $\alpha$-process, and the $\nu p$-process~\cite{Burbidge:1957vc,Cameron:1957PASP,Woosley:1992ek,Meyer:1994gn,Hoffman:1996ApJ,Frohlich:2005ys,Pruet:2005qd,Wanajo:2006ec,Fischer:2023ebq,Wang:2023tso}.

Note that the reactions in Eq.~\eqref{eq:CCnp}, and their efficacy in energy deposition and composition (isospin) determination, are flavor specific and neutrino or antineutrino specific.  Standard model neutrinos come in three weak interaction states, or flavors: $\nu_e$, $\nu_\mu$, $\nu_\tau$, with three corresponding antineutrino states: $\bar\nu_e$, $\bar\nu_\mu$, $\bar\nu_\tau$. In vacuum, these flavor states are related to the underlying mass eigenstates by a unitary transformation, which is described by the Pontecorvo–Maki–Nakagawa–Sakata matrix. The connection of neutrinos and antineutrinos with the mass basis can be different because of CP-violating phases: one for Dirac neutrinos and three for Majorana neutrinos. The way neutrinos couple in the weak interaction depends on their helicity, the projection of their spin along their propagation direction. In the low momentum transfer limit, left-handed neutrinos and right-handed antineutrinos couple weakly, while right-handed neutrinos and left-handed antineutrinos do not. The phenomenon we are interested in here, helicity conversion (which is also dubbed ``spin-flip''), has different consequences depending on the nature of neutrinos. Majorana neutrinos are those where spin-flip is tantamount to interchanging a neutrino and a corresponding antineutrino and {\it vice versa}, both \lq\lq active\rq\rq\ states that interact via the weak interaction. Spin-flip for Dirac neutrinos creates states that do not interact weakly. For Dirac neutrinos, conversion from a left-handed active state to a right-handed state can populate a sterile sector.  

Because the flavor and helicity composition of the neutrino field controls both energy deposition and lepton-number transport, neutrino quantum evolution in dense media is an essential part of the supernova problem. In addition to vacuum and matter-induced neutrino flavor transformation, neutrino--neutrino forward scattering gives rise to nonlinear collective flavor evolution~\cite{Duan:2010bg,Chakraborty:2016yeg,Tamborra:2020cul,Richers:2022zug,Volpe:2023met,Johns:2025mlm}. Most treatments of this problem use a mean-field or quantum-kinetic framework, where the many-body state is approximated by one-body density matrices evolving in a self-consistent potential~\cite{pantaleone:1992plb,Sigl:1993ctk,Volpe:2013uxl,Vlasenko:2013fja,Blaschke:2016xxt,Richers:2019grc,Froustey:2020mcq}. This approach has been very successful, but it discards any many-body correlations. In the flavor sector, exact many-body calculations and related studies have shown that entanglement, finite-number effects, and the structure of the interaction Hamiltonian can lead to behavior that is not always captured by mean-field approaches (see, e.g.,~\cite{Bell:2003mg,Pehlivan_2011,Rrapaj:2019pxz,Cervia:2019res,Roggero:2021asb,Patwardhan:2021rej,Xiong:2021evk,Martin:2021bri,Cervia:2022pro,Lacroix:2022krq,Amitrano:2022yyn,Illa:2022zgu,Martin:2023gbo,Turro:2024shh,Chernyshev:2024pqy,Siwach:2024jet,Spagnoli:2025etu,Bleau:2026iiq} and the reviews~\cite{Patwardhan:2022mxg,Balantekin:2023qvm})---see also, however, \cite{Friedland:2003dv, Friedland:2003eh, Friedland:2006ke, McKellar:2009py, Shalgar:2023ooi, Johns:2023ewj, Kost:2024esc, Goimil-Garcia:2024wgw, Kost:2025vyt} for discussions of several caveats that merit careful consideration in these situations. More recently, attention has turned to the role of non-forward terms in the full neutrino self-interaction Hamiltonian, which are absent from the usual mean-field potential but can connect distinct many-body momentum configurations~\cite{Cirigliano:2024pnm,Froustey:2026slw}.

Helicity conversion provides a closely related problem. 
Since neutrinos have nonzero mass, helicity is not exactly conserved for relativistic neutrinos. General quantum-kinetic treatments have shown that helicity coherence is sourced by terms proportional to $m/E_\nu$ (mass/energy) and by anisotropies in the medium, including transverse matter or neutrino currents~\cite{Cirigliano:2014aoa,Vlasenko:2014bva,Serreau:2014cfa,Kartavtsev:2015eva}. Such effects are expected to be small because of the $(m/E_\nu)^2$ suppression of wrong-helicity occupation, and usually require specific resonant conditions to occur~\cite{Vlasenko:2014bva,Chatelain:2016xva,Tian:2016hec,Purcell:2024bim,Fiorillo:2024wej}. We note that neutrino-antineutrino \emph{pair oscillations} are not necessarily suppressed by the neutrino mass, but their occurrence is debated~\cite{Huang:2026fcb,Fiorillo:2026byi}.

In the presence of magnetic fields, a nonzero neutrino magnetic moment (which can be unrelated to the neutrino mass in nonstandard models) can induce spin precession~\cite{Fujikawa:1980yx,Schechter:1981hw,Egorov:1999ah,deGouvea:2012hg,deGouvea:2013zp,Giunti:2014ixa,Ternov:2016njz,Dobrynina:2016rwy,Studenikin:2016oyh,Studenikin:2020mfh,Chukhnova:2019oum,Abbar:2020ggq,Chukhnova:2020xth,Giunti:2020dqw,Yuan:2021exm,Sasaki:2021bvu,Bulmus:2022gyz,Sasaki:2023sza,Manno:2026ikk}. However here again particularly large magnetic fields or neutrino magnetic moments are typically necessary to obtain a significant effect, with some exceptions. Nevertheless, even small changes in the active, sterile, or neutrino--antineutrino content could affect neutrino heating, lepton-number transport, the neutronization burst, or the electron fraction of the ejecta. We also note that spin-flip could lead to an observable difference between Dirac and Majorana neutrino scattering on electrons~\cite{Barranco:2014cda,Barranco:2017zeq}.

In this work, we study neutrino helicity conversion from a quantum many-body perspective. We consider a controlled toy model consisting of a finite set of momentum modes populated by one flavor of Dirac neutrinos of rest mass $m$. By restricting to a single flavor and initially left-handed neutrinos, we isolate helicity conversion from ordinary flavor transformation and antineutrino dynamics. We compare three descriptions of the same underlying physics: a mean-field evolution for helicity density matrices, an exact many-body evolution generated by a forward/exchange ``truncated'' Hamiltonian, and an exact many-body evolution generated by the full neutrino self-interaction Hamiltonian. The latter includes all momentum-exchange terms allowed by momentum conservation, with pairwise kinetic-energy conservation emerging dynamically from the hierarchy between the kinetic and interaction energy scales~\cite{Cirigliano:2024pnm}.

Our goal is not to provide a realistic supernova calculation, but to identify when and how many-body physics beyond the mean field can modify helicity conversion. We find that, in the one-flavor systems studied here, the truncated many-body calculation agrees with the mean-field result at small $m/E_\nu$. By contrast, the full many-body Hamiltonian opens additional momentum states through non-forward scattering, creating new channels for left-to-right conversion. In the finite systems explored below, this can enhance the total right-handed occupation by orders of magnitude, and in special symmetric configurations can produce particularly large-amplitude coherent oscillations of the helicity population.

This paper is organized as follows. In Sec.~\ref{sec:evolution_equations}, we derive the Hamiltonian for a system of one-flavor Dirac neutrinos at leading order in $m/E_\nu$, with technical details given in Appendix~\ref{app:QFT}. We then present the mean-field limit and define the full and truncated many-body Hamiltonians used in the numerical calculations. In Sec.~\ref{sec:results}, we compare mean-field, truncated many-body, and full many-body evolution for several finite momentum configurations. We introduce the setup for our calculations in Sec.~\ref{subsec:setup}, with algorithmic details in Appendix~\ref{app:numerical_details}. We discuss generic many-body enhancement in Sec.~\ref{subsec:general_results}. We connect our results with transverse-current mean-field behavior in Sec.~\ref{subsec:transverse_potential}. We study special large-conversion configurations in Sec.~\ref{subsec:large_conversion}. The scaling with $m/E_\nu$ is discussed in Appendix~\ref{app:m2_dependence}. We summarize our findings and discuss implications and future prospects in Sec.~\ref{sec:summary}.

Throughout this paper, we use natural units in which $\hbar=c=k_\mathrm{B}=1$.

\section{Neutrino evolution equations}
\label{sec:evolution_equations}

\subsection{Hamiltonian and Schrödinger equation}

\subsubsection{Expression of the Hamiltonian}

We focus here on the helicity transformation of a many-body system of neutrinos, and do not consider the flavor conversion more traditionally studied in the literature. We therefore restrict our study to a single flavor of Dirac neutrinos, with mass $m$. The Hamiltonian driving the evolution of a system of such interacting neutrinos is given by (see Appendix~\ref{app:interaction_picture} for notation and a rigorous derivation)
\begin{equation}
\label{eq:H_QFT}
\begin{aligned}
    H &= H_0 + H_{\nu \nu} \\
        &= \int{d^3 \vec{x} \, \bar{\nu}(\vec{x}) \left(- i \vec{\gamma} \cdot \vec{\nabla} + m\right)  \nu(\vec{x})}  \\ 
        &\qquad +  \frac{G_F}{\sqrt{2}} \int d^3 \vec{x} \, [\bar{\nu}(\vec{x}) \gamma^\sigma P_L \nu(\vec{x})] [\bar{\nu}(\vec{x}) \gamma_\sigma P_L \nu(\vec{x})] \, ,
\end{aligned}
\end{equation}
where $G_F\approx 1.166\times{10}^{-11}\, \mathrm{MeV}^{-2}$ is the Fermi constant.
We Fourier expand the fields on a discrete set of 3-momenta
\begin{equation}
\label{eq:def_p}
    \vec{p} = \frac{2 \pi}{L} \vec{n} \quad \text{with} \quad \vec{n} = \begin{pmatrix} n_x \\ n_y \\ n_z \end{pmatrix} \in \mathbb{Z}^3 \, ,
\end{equation}
such that the energy scale is given by $p_0 = 2 \pi/ L$, and the ``quantization volume'' is $V = L^3$. The fields are written
\begin{multline}
\label{eq:Fourier_expansion}
    \nu(\vec{x}) = \sum_{\vec{p}} \sum_{h = \pm}  \frac{1}{\sqrt{2VE_{\vec{p}}}}\bigg(a_h(\vec{p}) u_h(\vec{p}) e^{i \vec{p} \cdot \vec{x}} \\ 
    +
    b^\dagger_h(\vec{p}) v_h(\vec{p}) e^{- i \vec{p} \cdot \vec{x}} \bigg) \, ,
\end{multline}
with the expressions of the helicity spinors defined in Appendix~\ref{app:QFT}, Sec.~\ref{app:spinors}. The above creation and annihilation operators for left- and right-handed occupation obey the nonzero anticommutation rules
\begin{equation}
    \left\{ a_h(\vec{p}), a^\dagger_{h'}(\vpp)\right\} = \left\{ b_h(\vec{p}), b^\dagger_{h'}(\vpp)\right\} =\delta_{hh'} \delta_{\vec{p},\vpp} \, .
\end{equation}
Explicitly, the momentum delta-function is defined for $\vec{p}=(2 \pi / L) \vec{n}$ and $\vpp = (2 \pi / L) \vec{n}'$ as
\begin{equation}
    \delta_{\vec{p},\vpp}=\delta_{n_x n'_x} \delta_{n_yn'_y} \delta_{n_z n'_z} \, .
\end{equation}

With these definitions, we can write the second quantization version of the Hamiltonian in Eq.~\eqref{eq:H_QFT}. The vacuum component of this Hamiltonian is
\begin{equation}
    H_0 = \sum_{\vec{p},h} E_{\vec{p}} \, a^\dagger_h(\vec{p})a_h(\vec{p}) + \sum_{\vec{p},h} E_{\vec{p}} \, b^\dagger_h(\vec{p})b_h(\vec{p}) \, ,
\end{equation}
with $E_{\vec{p}} = \sqrt{{\vec{p}}^2 + m^2}$. For the interacting part of the Hamiltonian, we see that the general expression of $H_{\nu \nu}$ consists of $2^4 = 16$ terms, one for each neutrino/antineutrino operator combination. In particular, the term involving only neutrino operators is
\begin{equation}
\label{eq:Hnunu_1}
    H_{\nu\nu}^{(1)} = \mu \sum_{1\dots4} 
    \delta_{\vec{p}_1+\vec{p}_2,\vec{p}_3+\vec{p}_4} \, g(1,3,2,4) \, a^\dagger_1 a_3 a^\dagger_2 a_4 \, ,
\end{equation}
where $\mu \equiv G_F / (\sqrt{2}V)$,
\begin{equation}
    g(1,3,2,4) = \frac{ \left(\bar{u}_1 \gamma^\sigma P_L u_3 \right)\left(\bar{u}_2 \gamma_\sigma P_L u_4 \right) }{4\sqrt{E_{\vec{p}_1}E_{\vec{p}_2}E_{\vec{p}_3}E_{\vec{p}_4}}} \, ,
\end{equation}
and the indices $i = 1, \dots 4$ stand for both momentum $\vec{p}_i$ and helicity $h_i$. Note that the Fierz identities give 
\begin{equation}
    g(1,3,2,4) = - g(1,4,2,3) = - g(2,3,1,4).
\end{equation}

In this work, we restrict ourselves for simplicity to situations with only neutrinos. As a consequence, the Hamiltonian is $H = H_0 + H_{\nu\nu}^{(1)}$, the total particle number operator and the lepton number operator are identical:
\begin{equation}
    \mathrm{N} = \mathrm{L} = \sum_{\vec{p},h} a^\dagger_h(\vec{p}) a_h(\vec{p}) \, ,
\end{equation}
and the total momentum operator is
\begin{equation}
    \vec{P} = \sum_{\vec{p},h}\vec{p} \, a^\dagger_h(\vec{p}) a_h(\vec{p}) \, .
\end{equation}
One can easily verify that these operators correspond to conserved quantities, since $[H, \mathrm{N}] = 0$ and $[H, \vec{P}] = \vec{0}$.

The Schrödinger equation describing the evolution of the many-body quantum state $\ket{\Psi}$ is
\begin{equation}
\label{eq:Schrodinger}
    i \frac{\dd\ket{\Psi(t)}}{\dd t} = H\ket{\Psi(t)} = \left(H_0 + H_{\nu\nu}^{(1)}\right) \ket{\Psi(t)} \, .
\end{equation}
$H$ is independent of time, so that the solution of \eqref{eq:Schrodinger} is
\begin{equation}
\label{eq:time_evo}
    \ket{\Psi(t)} = e^{- i H t} \ket{\Psi(0)} \, ,
\end{equation}
and a simple approach to computing $e^{-i H t}$ can be obtained by diagonalizing $H$.

\subsubsection{Occupation numbers}

In the following, we will be particularly interested in the one-body observables describing the amount of right-handed neutrinos. The generalized occupation numbers are the combinations $\braket{ a^\dagger_{h'}(\vec{p}^{\, \prime}) a_{h}(\vec{p}) }$. The initial states we will consider will be product states $\ket{\Psi(0)} = a^\dagger_{\dots} \cdots a^\dagger_{\dots} \ket{0}$, which are eigenvectors of $\mathrm{N}=\mathrm{L}$ and $\vec{P}$. These operators commute with $H$ and therefore $e^{iHt}$, such that $\ket{\Psi(t)}$ is also an eigenstate of $\mathrm{N}=\mathrm{L}$ and $\vec{P}$, with the same eigenvalues. This leads to vanishing expectation values of the form $\bra{\Psi(t)}a^\dagger a^\dagger\ket{\Psi(t)}=\braket{a^\dagger a^\dagger}=0$ (same for a combination $aa$), and 
\begin{equation}
\label{eq:occ_number}
    \braket{ a^\dagger_{h'}(\vpp) a_{h}(\vec{p}) } \equiv \delta_{\vec{p},\vpp} \rho_{hh'}(\vec{p},t) \, ,
\end{equation}
where we introduce the momentum-diagonal part of the one-body density matrices
\begin{equation}
\label{eq:def_rho}
    \rho(\vec{p},t) = \begin{pmatrix} \rho_{--}(\vec{p},t) & \rho_{-+}(\vec{p},t) \\ \rho_{+-}(\vec{p},t) & \rho_{++}(\vec{p},t) \end{pmatrix} \, .
\end{equation}
For instance, the total occupation number of right-handed neutrino states is $\sum_{\vec{p}} \rho_{++}(\vec{p},t)$.

\subsection{Lowest-order description of helicity conversion}
\label{subsec:leading_order}

The typical neutrino energies in astrophysical environments are of order MeV, while the active neutrino masses are constrained below the eV scale. For instance, the KATRIN experiment sets the bound $m < 0.45 \, \mathrm{eV}$ at 90 \% confidence level based on precision spectroscopy of the tritium $\beta$-decay~\cite{KATRIN:2024cdt}.\footnote{Tighter constraints (sometimes close to being incompatible with lower bounds from oscillation experiments) are obtained in cosmology, using cosmic microwave background and large-scale structure measurements, although there are some tensions between datasets~\cite{DESI:2024hhd,DESI:2025zgx,SPT-3G:2025bzu,ACT:2025qjh,Craig:2024tky}.} Therefore, the mass effects are controlled by a very small parameter $\tilde{m} \equiv m/p_0 \ll 1$. We will concentrate on the leading order effects in $\tilde{m}$.

The leading contribution to the vacuum Hamiltonian $H_0$ is of zeroth order in $\tilde{m}$,
\begin{equation}
    H_0 = \sum_{\vec{p}}{p \left[ a^\dagger_+(\vec{p}) a_+(\vec{p}) + a^\dagger_-(\vec{p}) a_-(\vec{p}) \right]} + \mathcal{O}(\tilde{m}^2) \, .
\end{equation}
Writing $H^{(1)}_{\nu\nu}$ in normal ordering and keeping only the zeroth and first orders in $\tilde{m}$, we have
\begin{equation}
    H^{(1)}_{\nu\nu} \approx H_{LL \to LL} + H_{LL \to LR} + H_{LR \to LL} + \mathcal{O}(\tilde{m}^2) \, ,
\end{equation}
with
\begin{multline}
    H_{LL \to LL} = -\mu \sum_{\vec{p}_1\dots\vec{p}_4} \delta_{\vec{p}_1+\vec{p}_2,\vec{p}_3+\vec{p}_4} \, g(1_-, 3_-, 2_-, 4_-) \\ a^\dagger_-(\vec{p}_1) a^\dagger_-(\vec{p}_2)a_-(\vec{p}_3) a_-(\vec{p}_4) \, ,
\end{multline}
\begin{multline}
\label{eq:H_LL_LR}
    H_{LL \to LR} = - 2\mu \sum_{\vec{p}_1\dots\vec{p}_4} \delta_{\vec{p}_1 + \vec{p}_2 , \vec{p}_3+\vec{p}_4} \, g(1_+,3_-,2_-,4_-) \\
    a^\dagger_+(\vec{p}_1) a^\dagger_-(\vec{p}_2) a_-(\vec{p}_3) a_-(\vec{p}_4) \, ,
\end{multline}
and
\begin{multline}
\label{eq:H_LR_LL}
    H_{LR \to LL} = - 2\mu \sum_{\vec{p}_1\dots\vec{p}_4} \delta_{\vec{p}_1 + \vec{p}_2, \vec{p}_3 + \vec{p}_4} \, g(1_-,3_-,2_-,4_+) \\ a^\dagger_-(\vec{p}_1) a^\dagger_-(\vec{p}_2) a_-(\vec{p}_3) a_+(\vec{p}_4) \, . 
\end{multline}
The interaction matrix elements are
\begin{equation}
\begin{aligned}
    g(1_-,3_-,2_-,4_-) &= f^*(\hat{p}_1,\hat{p}_2) f(\hat{p}_3,\hat{p}_4) \, , \\
    g(1_+,3_-,2_-,4_-) &= \frac{m}{2p_1}h(\hat{p}_1,\hat{p}_2) f(\hat{p}_3,\hat{p}_4) \, , \\
    g(1_-,3_-,2_-,4_+) \,  &= -\frac{m}{2p_4}f^*(\hat{p}_1,\hat{p}_2) h(\hat{p}_3,\hat{p}_4) \, ,
\end{aligned}
\end{equation}
where $\hat{p} \equiv \vec{p}/|\vec{p}|$ is the unit vector in the direction of $\vec{p}$, and
\begin{widetext}
\begin{align}
    \label{eq:f_p_q}
    f(\hat{p},\hat{q}) &= \sqrt{2} \left[ e^{-i\phi_{\vec{p}}} \sin \left(\frac{\theta_{\vec{p}}}{2} \right)\cos \left(\frac{\theta_{\vec{q}}}{2}\right) - e^{-i\phi_{\vec{q}}} \cos \left(\frac{\theta_{\vec{p}}}{2}\right) \sin \left(\frac{\theta_{\vec{q}}}{2}\right)\right] \, , \\
    \label{eq:h_p_q}
    h(\hat{p},\hat{q}) &= \sqrt{2} \left[  \cos \left(\frac{\theta_{\vec{p}}}{2} \right) \cos \left(\frac{\theta_{\vec{q}}}{2}\right)+ e^{i\left(\phi_{\vec{q}}-\phi_{\vec{p}}\right)} \sin \left(\frac{\theta_{\vec{p}}}{2}\right)\sin \left( \frac{\theta_{\vec{q}}}{2}  \right)\right] \, ,
\end{align}
\end{widetext}
using the spherical coordinates for the momentum vectors, with angles $(\theta_{\vec{p}}, \phi_{\vec{p}})$. The expression of $f$, Eq.~\eqref{eq:f_p_q}, coincides with Eq.~(17) in~\cite{Cirigliano:2024pnm}, as expected for the left-left helicity part of the interaction Hamiltonian. 

Of particular interest in the literature is the \emph{forward limit} (see Sec.~\ref{subsec:full_trunc_H}), for which $\vec{p}_1=\vec{p}_3 = \vec{p}$ and $\vec{p}_2=\vec{p}_4 = \vec{q}$. In this case, the matrix elements read
\begin{widetext}
\begin{subequations}
\begin{align}
    g(1_-,1_-,2_-,2_-) &= 1 - \hat{p} \cdot \hat{q} \, , \label{eq:forward_matrix_element}   \\
    g(1_+,1_-,2_-,2_-) &=  \frac{m}{2p} e^{-i\phi_{\vec{p}}} \left[ \sin \theta_{\vec{p}} \cos \theta_{\vec{q}} - \cos \theta_{\vec{p}} \sin \theta_{\vec{q}} \cos\left(\phi_{\vec{p}} -\phi_{\vec{q}} \right) -i \sin\theta_{\vec{q}} \sin \left( \phi_{\vec{p}} -\phi_{\vec{q}} \right) \right] \, ,     \\ 
    g(1_-,1_-,2_-,2_+) &= \frac{m}{2q} e^{i\phi_{\vec{q}}} \left[ \cos \theta_{\vec{p}} \sin \theta_{\vec{q}} - \sin \theta_{\vec{p}} \cos \theta_{\vec{q}} \cos \left(\phi_{\vec{p}} -\phi_{\vec{q}}  \right) -i \sin \theta_{\vec{p}} \sin \left(\phi_{\vec{p}} - \phi_{\vec{q}} \right) \right] \, . \label{eq:g_---+}
\end{align}
\end{subequations}
\end{widetext}
We recognize in Eq.~\eqref{eq:forward_matrix_element} the usual strength of the left-left forward scattering~\cite{Duan:2010bg}. We note that all these matrix elements, in the forward limit, vanish if $\vec{p}=\vec{q}$.

\subsection{Mean-field approximation}

A common approach used in neutrino transport is the \emph{mean-field approximation}, where two-body (and higher order) correlations are discarded. In other words, the system is completely described by one-body quantities, namely, the one-body density matrix introduced in Eqs.~\eqref{eq:occ_number}--\eqref{eq:def_rho}. The evolution equation in this case can be obtained by applying the Ehrenfest theorem and reads~\cite{Volpe:2013uxl,Serreau:2014cfa} 
\begin{equation}
\label{eq:QKE}
    i\frac{\dd \rho(\vec{p},t)}{\dd t} = \left[ \Gamma(\vec{p},t), \rho(\vec{p},t) \right] \, ,
\end{equation}
with the mean-field potential
\begin{equation}
    \Gamma(\vec{p},t) = \begin{pmatrix} \Gamma_{--}(\vec{p},t) & \Gamma_{-+}(\vec{p},t) \\ \Gamma_{+-}(\vec{p},t) & \Gamma_{++}(\vec{p},t) \end{pmatrix} \, ,
\end{equation}
where
\begin{equation}
\label{eq:def_Gamma}
    \Gamma_{hh'}(\vec{p}_1,t) = 4\mu\sum_{\vec{p}_2, s, s'} g(1_h,1_{h'},2_s,2_{s'}) \rho_{s's}(\vec{p}_2,t) \, .
\end{equation}
Strictly speaking, there is also a term arising from the vacuum Hamiltonian in Eq.~\eqref{eq:QKE}, but it is proportional to the identity matrix and therefore does not contribute to the commutator. As can be seen in the expression in Eq.~\eqref{eq:def_Gamma}, only a subset of all two-body processes contribute at the mean-field level. Specifically, these are the ``forward'' and ``exchange'' processes, where the pair of initial and final momenta are identical: schematically, it corresponds to interactions $\nu_{\vec{p}_1} + \nu_{\vec{p}_2} \leftrightarrow \nu_{\vec{p}_1} + \nu_{\vec{p}_2}$, and $\nu_{\vec{p}_1} + \nu_{\vec{p}_2} \leftrightarrow \nu_{\vec{p}_2} + \nu_{\vec{p}_1}$.

Before simplifying the mean-field equations for the case we are interested in, we note that this is the lowest level approximation in terms of multiparticle correlations—since they are fully neglected. An order-by-order extension can be obtained, for instance, in the BBGKY-like hierarchy approach~\cite{Volpe:2013uxl}. In particular, making the additional assumptions of molecular chaos and that of a clear separation of scales between the duration of a scattering event and the macroscopic timescales of evolution, one can obtain a generalization of the Boltzmann equation with flavor mixing, the so-called quantum kinetic equation (QKE)~\cite{Froustey:2020mcq} (see, e.g.,~\cite{Sigl:1993ctk,Blaschke:2016xxt} for other derivations). In this framework, scattering processes other than the forward/exchange ones appear, but the associated timescale is the collision one, which is much larger than the Hamiltonian evolution timescale that drives the many-body evolution. This difference was recently studied for flavor conversion in~\cite{Froustey:2026slw}, and signals a conceptual disagreement between the assumptions underlying QKE and many-body studies~\cite{Shalgar:2023ooi,Johns:2023ewj}. Such considerations are outside the scope of this work, and we do not implement here intermediate frameworks between the mean-field and full many-body ones.

\subsubsection{Simplifications of the mean-field potential}

The solution to the mean-field equation is formally
\begin{equation}
    \rho(\vec{p},t) = U(\vec{p},t) \rho(\vec{p},0) U^\dagger(\vec{p},t) \, ,
\end{equation}
with
\begin{equation}
    U(\vec{p},t) = \mathcal{T} \exp{ \left( -i\int_{0}^{t} d\tau \Gamma(\vec{p},\tau) \right) } \, ,
\end{equation}
where $\mathcal{T}$ denotes a time ordering operator.
As a consequence of these expressions, if a momentum bin $\vec{p}$ is not initially occupied, its density matrix remains $\rho(\vec{p},t)=0$ for all times.

Similarly to our description of the many-body Hamiltonian in Sec.~\ref{subsec:leading_order}, we want to write the mean-field equation up to leading order (in the helicity conversion terms). The orders of the matrix elements of $\Gamma(\vec{p},t)$ in $\tilde{m}$ are determined by the $g$ factor and the density matrix in Eq.~\eqref{eq:def_Gamma}. The density matrix elements $\rho_{hh'}(\vec{p},t)$ can be expanded in non-negative integer powers of $\tilde{m}$, so the lowest order dependence of $\Gamma$ on $\tilde{m}$ can be bounded by the $g$ factor. This shows that we have at least $\Gamma_{-+} = \mathcal{O}(\tilde{m}^1)$ and $\Gamma_{++}=\mathcal{O}(\tilde{m}^2)$ from the expressions in Eqs.~\eqref{eq:forward_matrix_element}--\eqref{eq:g_---+}. Since the mean-field potential is itself dependent on the density matrix, we need to self-consistently determine the order of each entry. If neutrinos are strictly massless, only the $\Gamma_{--}$ entry survives and the mean-field equation for any $\vec{p}$ becomes
\begin{equation}
    i\frac{\dd}{\dd t} \begin{pmatrix}
        \rho_{--} & \rho_{-+} \\
        \rho_{+-} & \rho_{++}
    \end{pmatrix} \underset{\tilde{m} \, = \, 0}{=} \begin{pmatrix}
        0 & \Gamma_{--} \rho_{-+} \\
        -\Gamma_{--}\rho_{+-} & 0
    \end{pmatrix} \, .
\end{equation}
With an initial condition where $\rho_{-+}$ and $\rho_{+-}$ are $0$ at $t=0$, they remain so at all times. In the massive neutrino case, the lowest order in $\tilde{m}$ for the elements $\rho_{-+}(\vec{p},t)$ and $\rho_{+-}(\vec{p},t)$ is therefore at least $\tilde{m}^1$.

Let only momenta $\vec{p}_l$ with $l=1,\dots,N_p$ initially be occupied. The lowest order expansion of $\Gamma(\vec{p}_j,t)$ then reads
\begin{align}
    \label{eq:Gamma_LL}
    \Gamma_{--}(\vec{p}_j,t) &= 4\mu \sum_{l=1}^{N_p} \left(1-\hat{p}_j \cdot \hat{p}_l \right) \rho_{--}(\vec{p}_l,t) \, , \\
    \Gamma_{-+}(\vec{p}_j,t) &= \Gamma_{+-}^*(\vec{p}_j,t) \nonumber \\
    = 2 \mu\frac{m}{p_j} &e^{i\phi_j} \left(\hat{p}_{j,\theta} + i \hat{p}_{j,\phi} \right) \cdot \sum_{l=1}^{N_p} \hat{p}_l \,  \rho_{--}(\vec{p}_l,t) \, ,
    \label{eq:Gamma_LR}  \\
     =  2 \mu\frac{m}{p_j} &e^{i\phi_j} \left( \hat{p}_{j,\phi} - i \hat{p}_{j,\theta} \right) \cdot \sum_{l=1}^{N_p} \left( \hat{p}_j \times \hat{p}_l \right)  \rho_{--}(\vec{p}_l,t) \, , \nonumber \\
    \Gamma_{++}(\vec{p}_j,t) &= 0 \, ,  \label{eq:Gamma_RR} 
\end{align}
where we introduce the spherical coordinates for $\vec{p}_j = p_j \left(\sin \theta_j \cos \phi_j, \sin \theta_j \sin \phi_j, \cos \theta_j \right)$, and the unit vectors orthogonal to $\vec{p}_j$ are
\begin{equation}
\begin{aligned}
    \hat{p}_{j, \theta} &= \left(\cos \theta_j \cos \phi_j, \cos \theta_j \sin \phi_j, - \sin \theta_j\right) \, , \\
    \hat{p}_{j, \phi} &= \left(- \sin \phi_j , \cos \phi_j , 0 \right) \, .
\end{aligned}
\end{equation}
The expressions in Eqs.~\eqref{eq:Gamma_LL}--\eqref{eq:Gamma_RR} are in perfect agreement with the more general equations derived in Ref.~\cite{Serreau:2014cfa}. Specifically, our results correspond to a limited case of their Eqs.~(113)--(118),\footnote{To make the connection explicit, we recall that in our expressions $\mu = G_F/(\sqrt{2}V)$; for one flavor the trace terms in Ref.~\cite{Serreau:2014cfa} simply double the result; and the complex vector $\epsilon_q^*$ appearing in their Eq.~(114) reads $\hat{q}_\theta + i \hat{q}_\phi$.} written for one flavor, without a matter contribution and noting that, in our configurations where there is no initial $\rho_{-+} \neq 0$, an additional contribution to $\Gamma_{--}$ included in Ref.~\cite{Serreau:2014cfa} would be of order $\mathcal{O}(\tilde{m}^2)$. Our mean-field potential expressions are also in perfect agreement with, e.g., Ref.~\cite{Cirigliano:2014aoa}, specifically their Eqs.~(26)--(28) written for one mass generation and with only neutrinos in the medium.

The lowest order $\tilde{m}$ dependence of $\rho_{++}(\vec{p},t)$ is determined by solving Eq.~\eqref{eq:QKE} with Eqs.~\eqref{eq:Gamma_LL}--\eqref{eq:Gamma_RR}. The evolution of the element $\rho_{++}$ is given by
\begin{equation}
\label{eq:drho++_dt}
    i\frac{\dd \rho_{++}}{\dd t}  = 
        \Gamma_{+-} \rho_{-+} - \Gamma_{-+}\rho_{+-}.
\end{equation}
Each term on the right is at least of the order $\tilde{m}^2$, so $\rho_{++}(\vec{p},t)-\rho_{++}(\vec{p},0)$ is also at least $\mathcal{O}(\tilde{m}^2)$. 

\subsubsection{Two-dimensional systems}

In Sec.~\ref{sec:results}, we will focus on two-dimensional systems with $p_z = 0$, such that $\theta_{\vec{p}} = \pi/2$ for all $\vec{p}$. It is then possible to simplify Eq.~\eqref{eq:Gamma_LR} as
\begin{equation}
\label{eq:Gamma_LR_2D}
\begin{aligned}
    \Gamma_{-+}(\vec{p}_j) &= 2 \mu\frac{m}{p_j} \, i \, e^{i\phi_j} \sum_{l=1}^{N_p} \sin(\phi_l - \phi_j) \,  \rho_{--}(\vec{p}_l) \\
    &= 2 \mu\frac{m}{p_j} \left(- \hat{p}_{j,y} + i \hat{p}_{j,x}\right) \sum_{l=1}^{N_p} \left(\hat{p}_j \times \hat{p}_l\right)_z   \rho_{--}(\vec{p}_l) \, ,
\end{aligned}
\end{equation}
the second expression being more convenient to use on a Cartesian grid of momenta. Here we have omitted the time variable for brevity.

\subsection{Full and truncated Hamiltonians}
\label{subsec:full_trunc_H}

Of all the momentum pairs satisfying the conservation conditions $\vec{p}_1 +\vec{p}_2 = \vec{p}_3 + \vec{p}_4$, a special category is forward/exchange scattering, for which $\vec{p}_1=\vec{p}_3$ or $\vec{p}_1=\vec{p}_4$. These are the only processes known to contribute at the mean-field level (leading order in $G_F$). Because of this, most studies on many-body effects on flavor oscillations have also restricted the two-body Hamiltonian to these forward/exchange combinations.

Doing the same thing in the spin-flip case, the ``truncated Hamiltonian'' reads
\begin{multline}
\label{eq:Hnunu_truncated}
    H_{\nu\nu}^\mathrm{(1,fwd)} = - \mu \sum_{1,\dots,4} 
    \left(\delta_{\vec{p}_1, \vec{p}_3} \delta_{\vec{p}_2, \vec{p}_4} + \delta_{\vec{p}_1, \vec{p}_4} \delta_{\vec{p}_2, \vec{p}_3}\right) \\ g(1,3,2,4) \, a^\dagger_1  a^\dagger_2 a_3 a_4 \, .
\end{multline}
Note that, if $\vec{p}_3 = \vec{p}_4$, the forward scattering factor is $\delta_{\vec{p}_1,\vec{p}_3}\delta_{\vec{p}_2,\vec{p}_3}$, but
there is no double counting with the above expression since the $g$ factor is zero (see Sec.~\ref{subsec:leading_order}). 

We emphasize that, although the truncated many-body and mean-field approaches include the same forward/exchange processes, the former allows for multiparticle correlations beyond the one-body level, while they are explicitly neglected in the latter. In the flavor conversion case, this leads to significant differences in the evolution of one-body observables~\cite{Patwardhan:2022mxg,Balantekin:2023qvm}. For helicity conversion, we show in the next section (see in particular Sec.~\ref{subsec:general_results}) that because of the smallness of $\tilde{m}$, the many-body effects are actually negligible for our setup in the truncated case, contrary to the full Hamiltonian case.

\section{Spin oscillations: mean-field vs many-body}
\label{sec:results}

We now proceed to explore the features of neutrino evolution in the different setups introduced before: mean-field, truncated many-body and full many-body. Specifically, we will be particularly interested in the number of ``wrong helicity'' neutrino states that become occupied over time.

\subsection{Setup and pairwise kinetic energy conservation}
\label{subsec:setup}

\subsubsection{Momentum configuration}

For computational purposes, we introduce some dimensionless quantities: time will be measured in units of $t_0 \equiv \mu^{-1}$; and the unit of energy is $p_0 = 2 \pi/L$. We define
\begin{equation}
    \tilde{t} \equiv \frac{t}{t_0}  \, ,
\end{equation}
and recall that $\vec{n} = \vec{p}/p_0$ is a vector of integers; see Eq.~\eqref{eq:def_p}. In order to use realistic values, we take $p_0 \approx 1 \, \mathrm{MeV}$. The ``quantization volume'' is then
\begin{equation}
    V = L^3 = \left(\frac{2\pi}{p_0} \right)^3 \sim (1.2 \times 10^{-12} \, \mathrm{m})^3 \, ,
\end{equation}
and the unit of time is $t_0 = \sqrt{2} V / G_F \sim 20 \, \mathrm{ns}$. The authors of Ref.~\cite{Cirigliano:2024pnm} use a temperature and associated equilibrium density to set their values of $p_0$ and $V$, but the results are similar.

The Schrödinger equation can now be written in
terms of dimensionless operators:
\begin{equation}
\label{eq:Schrodinger_dimensionless}
    i \frac{\dd \ket{\Psi(\tilde{t})}}{\dd \tilde{t}} = \left(\tilde{H}_0+\tilde{H}^{(1)}_{\nu\nu}  \right)\ket{\Psi(\tilde{t})} \, ,
\end{equation}
with
\begin{equation}
\label{eq:Hamiltonian_dimensionless}
\begin{aligned}
     \tilde{H}_0 &= \frac{p_0}{\mu} \sum_{\vec{p} = p_0 \vec{n}}  |\vec{n}| \left(a^\dagger_+(\vec{p})a_+(\vec{p})+ a^\dagger_-(\vec{p})a_-(\vec{p})\right) \, , \\
     \tilde{H}^{(1)}_{\nu\nu} &= \mu^{-1} H^{(1)}_{\nu\nu} \, .
\end{aligned}
\end{equation}

\subsubsection{Kinetic energy conservation}
\label{subsec:kinetic_energy}

In the interaction Hamiltonian~\eqref{eq:Hnunu_1}, the only conservation law that explicitly appears is that of spatial momentum $\vec{p}_1 + \vec{p}_2 = \vec{p}_3 + \vec{p}_4$. There is no a priori pairwise kinetic energy conservation, $|\vec{p}_1| + |\vec{p}_2| = |\vec{p}_3| + |\vec{p}_4|$. Such an additional conservation law appears in the forward limit, where it is a simple consequence of the equality of incoming and outgoing momenta. It also appears in the QKE collision term~\cite{Sigl:1993ctk,Blaschke:2016xxt,Froustey:2020mcq}, because overall energy conservation before and after a collision reduces to kinetic energy conservation, since in the molecular chaos approximation incoming and outgoing particles are free and uncorrelated. For the full many-body problem evolution, as described in Eq.~\eqref{eq:Schrodinger}, it was shown in Ref.~\cite{Cirigliano:2024pnm} that kinetic energy is conserved, but \emph{dynamically}. This is a consequence of the hierarchy between the kinetic and two-body terms, which in our notations reads $p_0/\mu \gg 1$. Specifically, for the values quoted above, we have $p_0 / \mu \sim 10^{13}$. This hierarchy leads to a separation in the two-body Hamiltonian between ``blocks'' of given value of $|\vec{p}| + |\vec{q}|$, and the off-diagonal, kinetic energy non-conservation driving terms are negligible. We formalize this argument below using time-dependent perturbation theory.

Because of the hierarchy between the one-body and two-body terms, we write the quantum state at a given time as a superposition of the eigenstates of the ``unperturbed Hamiltonian'' $H_0$, that is
\begin{equation}
\label{eq:Psi_expan}
    \ket{\Psi(t)} = \sum_{n} c_n(t) e^{-iE_nt} \ket{n} \, ,
\end{equation}
where $\ket{n}$ are the eigenstates of $H_0$, with $H_0 \ket{n} = E_n \ket{n}$, and the coefficients $c_n(t)$ are time-dependent because of the ``perturbation'' $H_{\nu \nu}$. Those eigenstates are of the form $\ket{n} = \prod_{i \in \mathcal{P}_n} a^\dagger(\vec{p}_i)\, \ket{0}$, with the eigenvalue $E_n = \sum_{i \in \mathcal{P}_n} |\vec{p}_i|$. We write $\mathcal{P}_n$ to denote the set of momenta occupied in the eigenstate $\ket{n}$. Given our choice of initial state $\ket{\Psi(0)}= \ket{k} = \prod_{i \in \mathcal{P}_k} a^\dagger(\vec{p}_i) \ket{0}$, the first-order result for coefficients $c_n(t)$ is~\cite{LandauLifshitz}
\begin{align}
    c_n(t) &\approx \delta_{nk} - i \bra{n}H_{\nu\nu}^{(1)}\ket{k} \int_{0}^{t} dt' e^{-i\left(E_k-E_n\right)t'} \\ 
    &= \delta_{nk} + 
    \begin{cases} - i \, \bra{n}H_{\nu\nu}^{(1)}\ket{k} \, t & \text{if } E_n=E_k \, , \\
    \bra{n}{H_{\nu\nu}^{(1)}}\ket{k} \frac{e^{-i(E_k - E_n) t} - 1}{E_k - E_n} & \text{if } E_n \neq E_k \, . \end{cases} \nonumber
\end{align}
Recall that the time will be measured in units of $t_0 = \mu^{-1}$, $\mu$ being the typical scale of coefficients $\bra{n} H_{\nu \nu}^{(1)} \ket{k}$. The term for $E_n = E_k$ then gives an order 1 contribution, while if $E_n \neq E_k$ the contribution is suppressed by a factor $\propto \mu/p_0 \ll 1$. In other words, the sectors of different $E_n$, that is, of different total kinetic energy, are not dynamically connected.

As a consequence, we can explicitly enforce pairwise kinetic energy conservation in the interaction Hamiltonian, which has no impact on the dynamics but dramatically reduces the size of the Hilbert space, by removing unnecessary basis states that would never be dynamically explored:
\begin{multline}
\label{eq:Hnunu_kin}
    H_{\nu\nu}^{(1)} \simeq -\mu \sum_{1,\dots,4} \delta_{\vec{p}_1+\vec{p}_2, \vec{p}_3+\vec{p}_4} \,  \delta_{|\vec{p}_1| + |\vec{p}_2|, |\vec{p}_3|+|\vec{p}_4|}  \\ g(1,3,2,4) \, a^\dagger_1 a^\dagger_2 a_3 a_4 \, .
\end{multline}
With this two-body Hamiltonian, and for the initial states we consider, the quantum state is at all times an eigenstate of the vacuum Hamiltonian $H_0$. Therefore, the vacuum Hamiltonian has no effect on the dynamics, but simply sets a global phase. We keep it for completeness, noting that it would still be relevant if more flavors were included.

\subsubsection{Resolution strategy}

In the next section, we will compare the results obtained for the mean-field [Eq.~\eqref{eq:QKE}], as well as the truncated and the full many-body calculations [Eq.~\eqref{eq:Schrodinger_dimensionless}].

We always consider pure Hartree-Fock states as initial conditions:
\begin{equation}
\label{eq:psi_0}
    \ket{\Psi(0)} = \ket{\Psi_0} = \prod_{i \in \mathcal{S}_0} a^\dagger_{h_i}(\vec{p}_i) \ket{0} \, ,
\end{equation}
with $\mathcal{S}_0$ the set of initially occupied one-body states. This corresponds to the initial density matrices
\begin{equation}
\begin{aligned}
    \rho(\vec{p}_i,0) &= \begin{pmatrix} \delta_{h_i,-} & 0 \\ 0 & \delta_{h_i,+} \end{pmatrix} &&\text{if } i \in \mathcal{S}_0 \, ,\\
    &= 0 &&\text{otherwise.}
\end{aligned}
\end{equation}

For the mean-field case, if the initially occupied momentum modes are $\vec{p}_1,\dots,\vec{p}_{N_p}$, there are $N_p$ equations of the form~\eqref{eq:QKE} to be solved. The $N_p$ density matrices are converted to a $4N_p \times 1$ vector
\begin{equation}
    \bm{\rho}(\tilde{t}) =  \begin{pmatrix}
    \dots \\
        \rho_{--}(\vec{p}_i,\tilde{t}) \\
        \rho_{++}(\vec{p}_i, \tilde{t}) \\
        \Re \, \rho_{-+}(\vec{p}_i,\tilde{t}) \\
        \Im \, \rho_{-+}(\vec{p}_i,\tilde{t}) \\
        \dots
    \end{pmatrix} \, .
\end{equation}
We use the Scipy routine \texttt{solve\_ivp} to solve the system of $4 \times N_p$ equations written for the components of $\bm{\rho}$.

For the many-body case, we adapted the code from Ref.~\cite{Cirigliano:2024pnm}, following the exact same strategy. We first determine, for a given initial state, the set of available states by applying $\tilde{H}_{\nu\nu}^{(1)}$, with increasing integer powers, to $\ket{\Psi_0}$, until no new states are explored (see Appendix~\ref{app:numerical_details}). This provides a basis for the Hilbert space in the form $\left\{ \ket{v_\alpha} \right\}_{\alpha = 1, \dots, k}$, where each $\ket{v_\alpha}$ is a pure Hartree-Fock state as in Eq.~\eqref{eq:psi_0}. We then have a basis in which we diagonalize the Hamiltonian $\tilde{H}$,
\begin{equation}
\label{eq:diagonalization_H}
    \tilde{H} \ket{\lambda_\alpha} = \lambda_\alpha \ket{\lambda_\alpha} \, ,
\end{equation}
obtaining the eigenstates $\ket{\lambda_\alpha}$ as a function of the basis states $\ket{v_\beta}$. The quantum state at time $\tilde{t}$ is then
\begin{equation}
\label{eq:psi_t}
    \ket{\Psi(\tilde{t})} = \sum_{\alpha}  \braket{\lambda_\alpha | \Psi(0)} \, e^{-i\lambda_\alpha \tilde{t}} \, \ket{\lambda_\alpha} \, .
\end{equation}
We are particularly interested in the number of right-handed neutrinos for a given momentum mode, $\rho_{++}$. While this is directly a variable in the mean-field case, in the many-body case it is obtained from the quantum state~\eqref{eq:psi_t} as
\begin{align}
    \rho_{++}(\vec{p},\tilde{t}) &= \braket{\Psi(\tilde{t}) | a^\dagger_+(\vec{p}) a_+(\vec{p}) | \Psi(\tilde{t})} \nonumber
    \\ &= \sum_{\alpha} r_{\alpha\alpha} (\vec{p}) + \sum_{\alpha\neq\beta} r_{\alpha\beta}(\vec{p})  e^{i\left(\lambda_\alpha -\lambda_\beta \right) \tilde{t}} \, , \label{eq:rho++}
\end{align}
with
\begin{equation}
    r_{\alpha\beta}(\vec{p}) = \braket{\Psi_0|\lambda_\alpha} \braket{\lambda_\beta|\Psi_0}  \braket{\lambda_\alpha| a^\dagger_+(\vec{p}) a_+(\vec{p}) |\lambda_\beta} \, . 
\end{equation}
Note that Eq.~\eqref{eq:rho++} yields a real number, since the $(\alpha, \beta)$ and $(\beta, \alpha)$ contributions are complex conjugates and these are summed.

Formally, it would be possible to extend our framework to \emph{mixed} many-body states, for which the full $N$-body density matrix could be represented as a classical admixture of multiple many-body pure states, i.e., $\hat D = \sum_j P_j \ket{\Psi}_j \bra{\Psi}_j$. Here, each $\ket{\Psi}_j$ is a many-body pure state, and the $P_j$'s are classical probabilities adding up to unity. The evolution of the system would then be described by a Liouville-von Neumann equation with the same many-body Hamiltonian that appears in our Schr\"odinger equation. The one-body observables in this case could be obtained as, e.g., $\rho_{++}(\vec{p},\tilde{t}) = \Tr\left[\hat D(\tilde t) \, a^\dagger_+(\vec{p}) a_+(\vec{p})\right]$. Such computations would require some structural changes in our numerical code, and we leave such considerations for future work.

\subsection{Enhancement of helicity conversion by many-body effects}
\label{subsec:general_results}

In the following, we will restrict ourselves to a two-dimensional grid of momenta:
\begin{equation}
    \label{eq:2D_grid}
    \vec{n} = \{ (n_x, n_y); \ n_x,n_y \in \mathbb{Z}   \} \ \ \text{with} \ \ 0< |n_x|, |n_y| \leq n_\text{max} .
\end{equation}
There is no grid point at $\vec{n}=(0,0)$ because all neutrinos are moving ultra-relativistically. The selection of values for $n_\text{max}$ depends on which momentum modes are initially occupied. One example, with the detailed numerical implementation of the construction and diagonalization of the Hamiltonian, is outlined in Appendix~\ref{app:numerical_details}. Furthermore, the initial state will always be composed of left-handed neutrinos only, i.e., in Eq.~\eqref{eq:psi_0} all $h_i = -$. This choice of initial state could mimic the realistic population hierarchy in a supernova-like environment, e.g., at the neutronization pulse where we would expect a small population of right-handed neutrinos.

In this section, we describe two randomly picked test cases, which highlight our main result: \emph{many-body effects significantly boost left-to-right helicity conversion, provided the full Hamiltonian is utilized}.

\begin{figure}[!ht]
    \centering
    \includegraphics[width=\columnwidth]{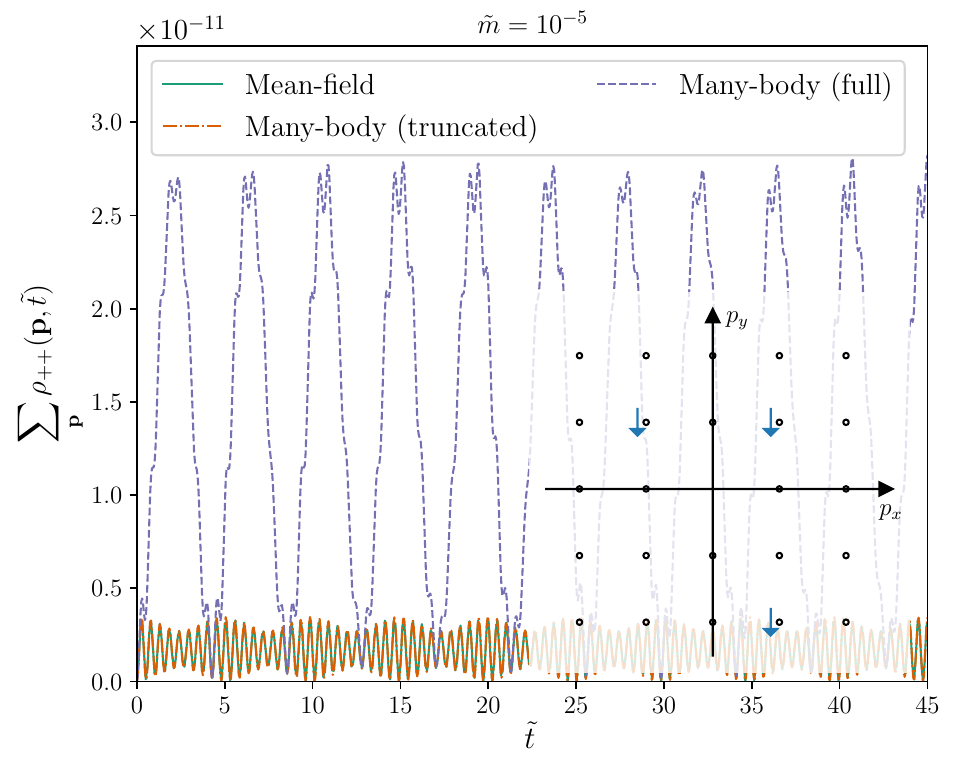}
    \caption{Evolution of the total occupation number of right-handed neutrino states on the two-dimensional momentum grid with $n_\text{max}=2$, modeled by the mean-field (solid green), truncated many-body (dash-dotted orange) and exact many-body (dashed purple) equations, for $\tilde{m}=10^{-5}$. The initial configuration, shown in the lower right corner, has three left-handed (blue downarrow) neutrinos with momenta $\vec{n}_3=(1,-2)$, $\vec{n}_{15}=(-1,1)$, and $\vec{n}_{17}=(1,1)$.}
    \label{fig:timeevolve_allleft}
\end{figure}

\begin{figure}[!ht]
    \centering
    \includegraphics[width=\columnwidth]{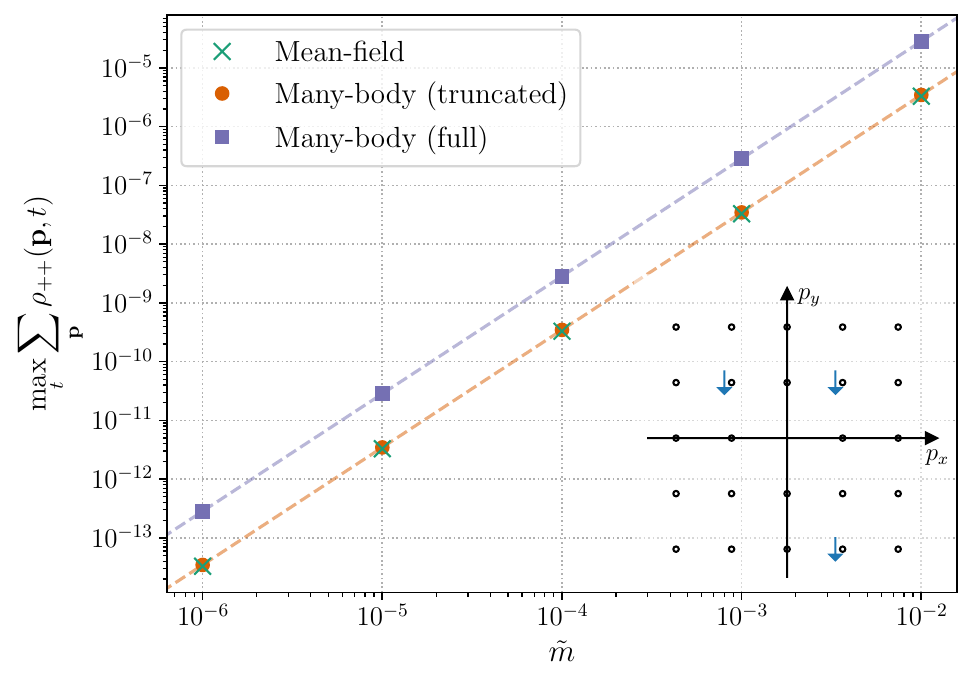}
    \caption{Maximum total right-handed occupation number (for $\tilde{t} \leq 100$), starting from the same configuration as in Fig.~\ref{fig:timeevolve_allleft}. We compare the results for the mean-field (green crosses), truncated many-body (orange dots) and full many-body (purple squares) calculations, for different values of the dimensionless mass parameter $\tilde{m} = m/p_0$. The dashed lines correspond to the $\propto \tilde{m}^2$ scaling of the left-to-right helicity conversion rate.}
    \label{fig:loglog}
\end{figure}

We first consider three left-handed neutrinos in the momentum bins $(1,-2)$, $(1,1)$, and $(-1,1)$. The evolution of the total right-handed occupation number, $\sum_{\vec{p}} \rho_{++}(\vec{p})$, is shown as a function of time in Fig.~\ref{fig:timeevolve_allleft} for a value of the dimensionless mass parameter $\tilde{m} = 10^{-5}$. 
This is larger than a realistic value ($\tilde{m} < 10^{-7}$), but we use it to avoid numerical problems in the resolution of the mean-field differential equation (which would require extremely small solver tolerances). We can thus illustrate the main features of spin-flip in the different cases we consider, confidently so given the extrapolation of the results to other values of $\tilde{m}$ shown in Fig.~\ref{fig:loglog}.
In that figure, we represent the maximum value of the total right-handed occupation number, which is a measure of the maximum spin-flip effect in our calculations. This quantity scales as $\propto \tilde{m}^2$ in all models, as expected in the mean-field case [see discussion around Eq.~\eqref{eq:drho++_dt}], but also in the many-body case as explained in Appendix~\ref{app:m2_dependence}.  For visualization purposes, the momentum grid with the initially occupied states is displayed in the bottom right of Figs.~\ref{fig:timeevolve_allleft} and \ref{fig:loglog}.

The amplitude of the total right-handed population in the exact many-body calculation is about $10$-times larger than in the other two models. Figure~\ref{fig:loglog} shows that this $10$-times enhancement is independent of $\tilde{m}$. We attribute this difference to the ``opening'' of new momentum states by the non-forward terms in the full Hamiltonian, which enables more left-to-right conversion channels.  Furthermore, the mean-field and truncated many-body calculations are indistinguishable, an intriguing feature compared with the study of flavor oscillations, which usually show differences between two such calculations (see, e.g.,~\cite{Rrapaj:2019pxz,Cervia:2019res,Roggero:2021asb,Patwardhan:2021rej}). We claim that this identicalness is a consequence of the specific form of the interaction Hamiltonian, with only one flavor present, and of the smallness of $\tilde{m}$. Indeed, we first note that the left-handed occupation number is almost unchanged (the difference scaling as $\tilde{m}^2$). As a consequence, the quantum state at all times is essentially an eigenvector of the occupation number operators $a^\dagger_-(\vec{p}_i)a_-(\vec{p}_i)$, which can, up to a correction $\propto \tilde{m}^2$, be replaced by the identity $\mathbb{I}$ if $i \in \mathcal{S}_0$, and $0$ otherwise. Since the truncated interaction Hamiltonian~\eqref{eq:Hnunu_truncated} is precisely made of products of such operators,\footnote{This would not be true anymore had we considered more than one flavor of neutrino. In that case, we would generally see differences between the forward many-body and mean-field calculations, as in the standard flavor oscillation literature.} it effectively reduces to a one-body term composed of $a^\dagger_{\mp}(\vec{q})a_{\pm}(\vec{q})$ combinations. For a Hamiltonian made of one-body operators, the mean-field approximation is exact, which justifies the identity of the mean-field and truncated many-body results. We have numerically checked that using unphysical, $\mathcal{O}(1)$ values for $\tilde{m}$, breaks this degeneracy between the mean-field and truncated calculation results, but in the interest of brevity we do not show it here. In what follows, we will not show the redundant results of mean-field calculations. Note that the above argument breaks down for the full Hamiltonian case, since the non-forward/exchange terms cannot be written as products of number operators for a given momentum.

\begin{figure}[!ht]
    \centering
    \includegraphics[width=\columnwidth]{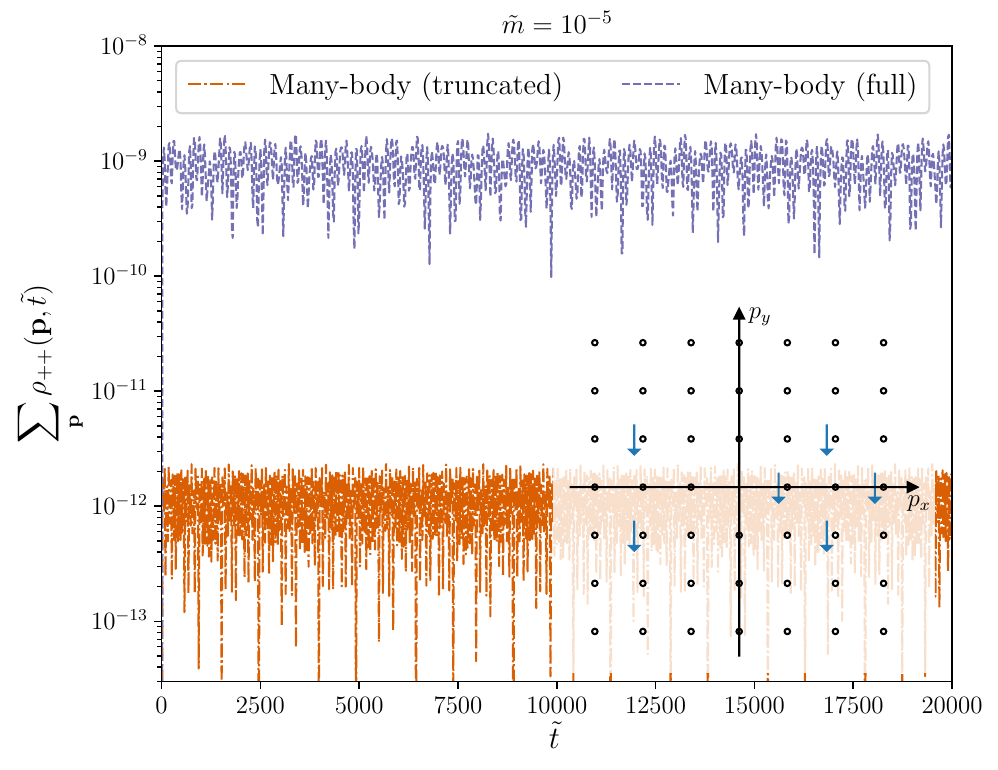}
    \includegraphics[width=\columnwidth]{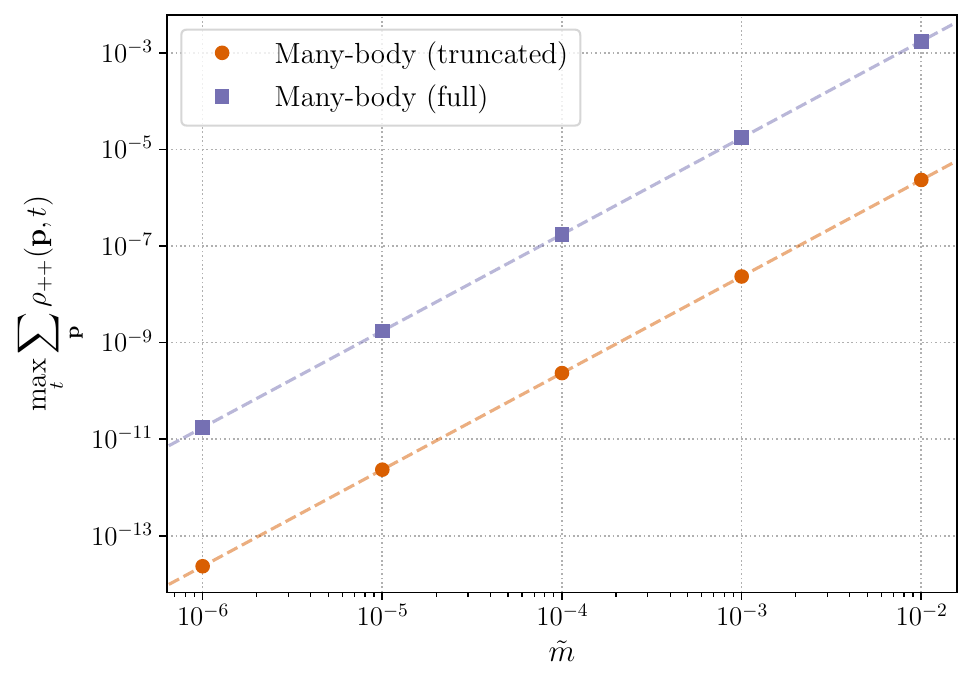}
    \caption{Same as Figs.~\ref{fig:timeevolve_allleft} and \ref{fig:loglog} for a different initial configuration (the mean-field results, identical to the truncated many-body ones, are not shown). The enhancement of helicity conversion in the full many-body case is even larger in this case.}
    \label{fig:example_twobeams}
\end{figure}

We consider in Fig.~\ref{fig:example_twobeams} a second example, where the initial state resembles an asymmetric ``beam-like'' configuration, with 4 neutrinos close to the $+ \hat{x}$ direction, and 2 neutrinos close to the $- \hat{x}$ direction. The top panel is similar to Fig.~\ref{fig:timeevolve_allleft}, but in log scale to make visible both the full and truncated many-body results. Compared to the previous example, there is an even larger, $10^3$-enhancement of the left-to-right conversion by the exact many-body model. The bottom panel of Fig.~\ref{fig:example_twobeams} shows, similarly to Fig.~\ref{fig:loglog}, the $\tilde{m}^2$ scaling of the right-handed maximal population, and the $\tilde{m}$-independent enhancement by the full-Hamiltonian's many-body effects.

This initial configuration once again allows for more final momentum states to be accessed by non-forward scatterings. On the other hand, if the initial configuration is such that, for a given momentum grid, there are very few differences between the accessible one-body states via the truncated and full Hamiltonians, then the left-to-right conversion enhancement will vanish. Numerically speaking, if the number of basis states $\{\ket{v_\alpha} \}_{\alpha=1,\dots,k}$ [see discussion before Eq.~\eqref{eq:diagonalization_H}] is much larger in the full many-body case than in the truncated case, the increased conversion is more likely. This picture is confirmed in the next subsection by employing a controlled test case.

\subsection{Transverse current and helicity conversion}
\label{subsec:transverse_potential}

At the mean-field level, the expressions~\eqref{eq:Gamma_LL}--\eqref{eq:Gamma_RR} show that the potential responsible for helicity conversion ($\Gamma_{-+} \propto \hat{p} \times \hat{q}$) is maximal for neutrinos propagating in transverse directions. This is a well-known result in the mean-field literature: a transverse (neutrino or matter) current is needed to build-up helicity coherence (see, e.g.,~\cite{Vlasenko:2013fja,Serreau:2014cfa, Kartavtsev:2015eva, Chatelain:2016xva, Purcell:2024bim}). 

\begin{figure}[!ht]
    \centering
    \includegraphics[width=0.7\columnwidth]{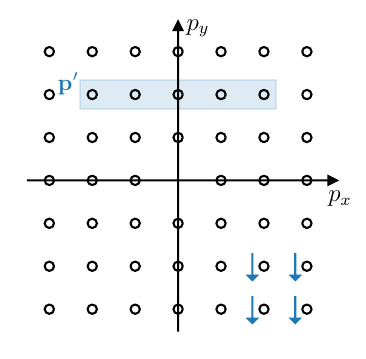}
    \caption{Initial configurations considered for the calculations in Sec.~\ref{subsec:transverse_potential}. There are initially 5 occupied momentum states with left-handed neutrinos: 4 in the bottom right corner, on or close to the direction $(+1,-1)$, and a fifth one that will change from calculation to calculation. This fifth momentum state, denoted $\vec{p}'$, is located in the blue band, and has coordinates $(n'_x \in [-2,2], n'_y = 2)$.}
    \label{fig:varying_angle_setup}
\end{figure}

This can be understood classically, as a potential (or force) transverse to the particle's propagation and spin direction will produce a torque and cause the spin to precess.  
In order to verify that we have the same phenomenology in our neutrino gas system, and how it changes in the full many-body treatment, we consider the configuration shown on Fig.~\ref{fig:varying_angle_setup}.

The system is composed of a majority of neutrinos traveling around the $\phi = - 45^\circ$ direction, and an additional neutrino with $p'_y > 0$. We will dub this latter neutrino the ``test'' neutrino $\vec{p}' = p_0 \vec{n}'$, since we will focus on the helicity transformation in that momentum bin. That transformation will stem from the interaction of the test neutrino with the other ``background'' neutrinos. We emphasize that all neutrinos are actually equivalent in this closed 5-body quantum system, but we first focus on one to highlight the mechanisms of spin flip. We will consider five different initial configurations, where $p'_x$ varies from $-2 p_0$ to $+ 2 p_0$, which allows us to probe different angles between the direction of propagation of this test neutrino, and the direction of the  background neutrino flux.

\subsubsection{Mean-field behavior}

Let us focus on the density matrix describing the $\vec{p}'$ momentum bin, first in the mean-field scenario.

We can make several simplifying assumptions: given the very small amount of helicity transformation (suppressed by $\tilde{m}^2$), we can take $\rho_{--}$ to be constant in time, equal to $1$ for initially occupied momentum states, and $0$ otherwise. As a consequence, the mean-field potentials at leading order, Eqs.~\eqref{eq:Gamma_LL}--\eqref{eq:Gamma_RR}, are also constant in time.

The left-right component of the mean-field equation~\eqref{eq:QKE} reads
\begin{equation}
    i \frac{\dd \rho_{-+}}{\dd t} = \Gamma_{--} \rho_{- +} - \rho_{--} \Gamma_{- +} \, ,
\end{equation}
where we omit the momentum argument in all quantities. The solution, considering $\rho_{-+}(t=0) = 0$ and the constancy of the different quantities, reads
\begin{equation}
\label{eq:rho_LR_example}
    \rho_{-+}(t) = \frac{\Gamma_{-+}}{\Gamma_{--}} \rho_{--} \left(1 - e^{- i \, \Gamma_{--} t}\right) \, .
\end{equation}
Turning now to the right-right component of the mean-field equation, we have
\begin{equation}
    i \frac{\dd \rho_{++}}{\dd t} = \Gamma_{+-} \rho_{- +} - \Gamma_{- +} \rho_{+-} = 2 \, i \, \Im(\Gamma_{+ -} \rho_{-+}) \, .
\end{equation}
Note that by trace conservation, $\dot{\rho}_{--} = - \dot{\rho}_{++}$, but the former is negligible compared to $\rho_{--}(0) = 1$, but not negligible compared to the latter. Inserting the expression~\eqref{eq:rho_LR_example}, and using $\Gamma_{+-} \Gamma_{-+} = |\Gamma_{-+}|^2$ and $\Gamma_{--} \in \mathbb{R}$, we can solve for
\begin{equation}
    \rho_{++}(t) \approx 2 \rho_{--} \left\lvert \frac{\Gamma_{-+}}{\Gamma_{--}}\right\rvert^2 \left[1 - \cos(\Gamma_{--}t)\right] \, .
\end{equation}
In particular, since $\rho_{--} \approx 1$, the maximum value of $\rho_{++}(t)$ is $4 |\Gamma_{-+}/\Gamma_{--}|^2$. If we now focus on the ``test'' momentum $\vec{p}'$, the ratio of Eqs.~\eqref{eq:Gamma_LR_2D} and \eqref{eq:Gamma_LL} gives
\begin{equation}
\begin{aligned}
    \left\lvert \frac{\Gamma_{-+}}{\Gamma_{--}}\right\rvert &= \frac12 \frac{m}{p'} \frac{\left \lvert \sum_{\vec{p} \neq \vec{p}'} \sin(\phi_{\vec{p}} - \phi_{\vec{p}'}) \rho_{--}(\vec{p}) \right \rvert}{\left \lvert \sum_{\vec{p} \neq \vec{p}'} [1-\cos(\phi_{\vec{p}} - \phi_{\vec{p}'})] \rho_{--}(\vec{p}) \right \rvert} \\
    &\approx \frac12 \frac{m}{p'} \left \lvert \frac{\sin(\Delta \phi)}{1 - \cos(\Delta \phi)}\right \rvert \, ,
\end{aligned}
\end{equation}
where we considered that the four ``background'' neutrinos, for which $\rho_{--} \simeq 1$, are roughly in the same direction $(+1, -1)$, and $\Delta \phi \in [\pi/2, \pi]$ is the angle between this direction and the direction of $\vec{p}'$. We finally get the prediction for the maximum amount of helicity conversion in the test momentum bin as a function of its direction:
\begin{equation}
    \label{eq:MF_prediction}
    \max_{t}{\rho_{++}(\vec{p}', t)} \approx \left(\frac{m}{p'}\right)^2 \cot^2\left(\frac{\Delta \phi}{2}\right) \, .
\end{equation}
As expected from the ``transverse potential'' picture, this quantity is maximal for $\Delta \phi = \pi/2$, and vanishes for $\Delta \phi = \pi$.

\begin{figure}[!ht]
    \centering
    \includegraphics[width=0.9\columnwidth]{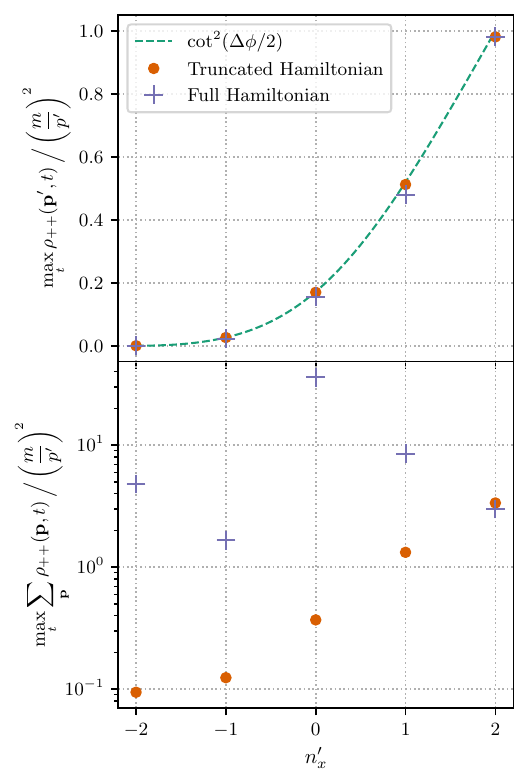}
    \caption{\emph{Top:} maximum right-handed helicity occupation number in the ``test'' momentum bin, in many-body calculations with the forward (orange dots) and full (purple crosses) Hamiltonians, for the different initial configurations shown on Fig.~\ref{fig:varying_angle_setup}. The occupation number is normalized by $(m/|\vec{p}'|)^2$, which is different for each calculation. The green dashed line is the mean-field prediction in Eq.~\eqref{eq:MF_prediction}, in perfect agreement with the truncated Hamiltonian calculation. \emph{Bottom:} maximum \emph{total} right-handed helicity occupation number, showing the enhancement of helicity conversion when using the full many-body Hamiltonian. We use the same normalization factor as in the top panel.}
    \label{fig:varying_angle_results}
\end{figure}

We show the maximum occupation number of the right-helicity state for the test momentum in the top panel of Fig.~\ref{fig:varying_angle_results}, comparing results obtained in many-body calculations with the truncated or full Hamiltonians. The truncated Hamiltonian results agree perfectly with the mean-field prediction~\eqref{eq:MF_prediction}, as expected from the general agreement between mean-field and forward many-body results shown in Sec.~\ref{subsec:general_results}. This confirms that our simulation reproduces the known behavior of spin transformation obtained in other mean-field studies. The full Hamiltonian result is very similar, with a slightly reduced amount of right-handedness. However, when looking at the \emph{total} right-handed helicity population (bottom panel of Fig.~\ref{fig:varying_angle_results}), we recover the enhancement of spin conversion discussed previously.

\begin{figure*}
    \centering
    \includegraphics[width=\textwidth]{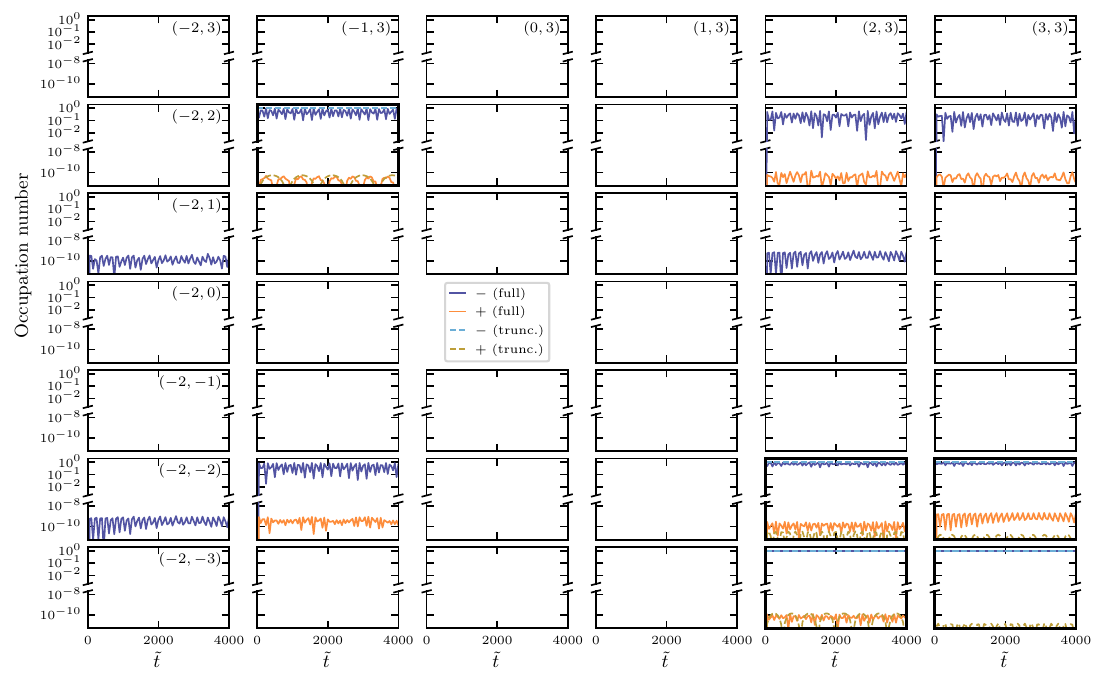}
    \caption{Time evolution of the occupation numbers for the initial configuration shown in Fig.~\ref{fig:varying_angle_setup}, with $\vec{n}' = (-1,2)$. Each panel is located on the grid location of the associated momentum, and we do not display the leftmost part of the grid ($n_x = -3$) since no bin is ever occupied. The initially occupied momentum bins are highlighted by thicker panel edges. The dashed (solid) lines show the results obtained with the truncated (full) Hamiltonian, with blue tones for the left-helicity and orange tones for the right-helicity quantities.}
    \label{fig:varying_angle_grid1}
\end{figure*}

\subsubsection{Accessible momentum states and helicity conversion}

This example sheds light on the mechanisms behind the increased spin flip observed when using the full Hamiltonian. We have argued previously that, since the general Hamiltonian allows for the population of new momentum states, these extra possibilities create new conversion channels and hence, more right-handed neutrinos. This can be checked concretely on the examples in Fig.~\ref{fig:varying_angle_setup}.

\begin{figure}[!ht]
    \centering
    \includegraphics[width=\columnwidth]{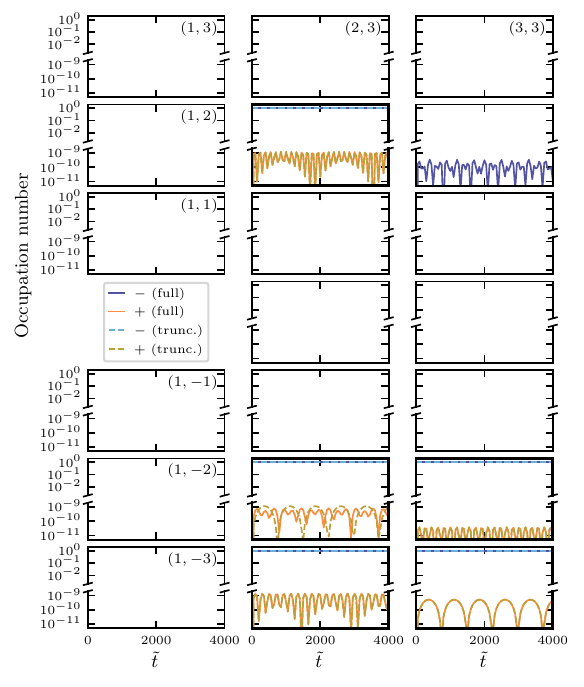}
    \caption{Same as Fig.~\ref{fig:varying_angle_grid1}, but with the fifth initial momentum state $\vec{n}' = (2,2)$. The momentum bins with $p_x \leq 0$ are never occupied and therefore not shown here.}
    \label{fig:varying_angle_grid2}
\end{figure}

In particular, we see in the bottom panel of Fig.~\ref{fig:varying_angle_results} that we generally have an enhancement of spin conversion, except in the $\vec{n}' = (2,2)$ case. We compare this scenario with the $\vec{n}' = (-1,2)$ case in Figs.~\ref{fig:varying_angle_grid1} and \ref{fig:varying_angle_grid2}.

Figure~\ref{fig:varying_angle_grid1} shows that several other momentum bins are being occupied through the action of the full Hamiltonian $H_{\nu \nu}$. For instance, starting from the initial configuration, the following energy and momentum conserving processes are possible (we describe the momenta by their integer coordinates in units of $p_0$):
\begin{equation}
    \begin{pmatrix} -1 \\ 2 \end{pmatrix}_- + \begin{pmatrix} 2 \\ -2 \end{pmatrix}_- \longrightarrow \begin{pmatrix} -1 \\ -2 \end{pmatrix}_\pm + \begin{pmatrix} 2 \\ 2 \end{pmatrix}_\mp \, ,
\end{equation}
increasing the average occupation number of right-handed states. The same exchange is possible between left-helicity states only, which leads to the high occupation of the bins $(-1,-2)_-$ and $(2,2)_-$. The same argument applies to occupy the $(3,2)$ state.

We also note that there are three momentum bins where a small population of left-handed state appears: $(-2,1)$, $(-2,-2)$, and $(2,1)$. For example, the first momentum bin gets populated via the process
\begin{equation}
\label{eq:secondary_process}
    \begin{pmatrix} -1 \\ 2 \end{pmatrix}_{h} + \begin{pmatrix} 2 \\ -3 \end{pmatrix}_{h'} \longrightarrow \begin{pmatrix} 3 \\ -2 \end{pmatrix}_{h''} + \begin{pmatrix} -2 \\ 1 \end{pmatrix}_- \, ,
\end{equation}
where one of the first three helicities must be $+$, which explains the very small amplitude of this process. The same momentum exchange with all left helicities ($h=h'=h''=-$) is forbidden by Pauli blocking, since in the initial configuration these three momentum bins are occupied by a left-handed neutrino. Overall, the extra-accessible momentum states allow for more helicity conversion than in the forward case.

This is different in the configuration of Fig.~\ref{fig:varying_angle_grid2}. In addition to the initially occupied momentum bins, there is only one other that can be populated, namely, $(3,2)$. Furthermore, the only process populating that bin is
\begin{equation}
    \begin{pmatrix} 2 \\ 2 \end{pmatrix}_{h} + \begin{pmatrix} 3 \\ -2 \end{pmatrix}_{h'} \longrightarrow \begin{pmatrix} 2 \\ -2 \end{pmatrix}_{h''} + \begin{pmatrix} 3 \\ 2 \end{pmatrix}_- \, ,
\end{equation}
where one of $h$, $h'$ or $h''$ is a right helicity, for the same reason as in Eq.~\eqref{eq:secondary_process}. Since this is the only difference with the forward case, it leads to a negligible variation of the total number of right-handed neutrinos, see Fig.~\ref{fig:varying_angle_results}.

\subsection{Large spin-conversion configuration}
\label{subsec:large_conversion}

In addition to the general many-body enhancement of spin conversion observed in the various previous examples, there are specific configurations which lead to an even starker enhancement. They correspond to situations where the total number of right-handed neutrinos appears to evolve with time as a cosine-like function with large amplitude and/or period. This is illustrated in the following subsections; see, in particular, Figs.~\ref{fig:compare_frequencies} and \ref{fig:amplitude_m2_compare_frequencies}.

In order to shed light on this behavior, we sum Eq.~\eqref{eq:rho++} over momentum modes to get
\begin{equation} 
\label{eq:sumrhopp}
    \sum_{\vec{p}} \rho_{++}(\vec{p},\tilde{t}) = \sum_{\alpha} R_{\alpha\alpha} +  \sum_{\alpha \neq \beta}  R_{\alpha\beta} e^{i \left(\lambda_\alpha - \lambda_\beta\right) \tilde{t}} \, ,
\end{equation}
with 
\begin{equation}
    R_{\alpha\beta} = \sum_{\vec{p}} r_{\alpha\beta}(\vec{p}) \, .
\end{equation}
If there are two integers $\mu$ and $\nu$, in $\{1,\ldots,k\}$ with $\mu \neq \nu$ such that $R_{\mu\nu}$ and $R_{\nu\mu}$ are much bigger than the other $R_{\alpha\beta}$ on the right-hand side of Eq.~\eqref{eq:sumrhopp}, the equation can be approximated as 
\begin{multline}
    \sum_{\vec{p}} \rho_{++}(\vec{p},\tilde{t}) \approx  \sum_{\alpha} R_{\alpha\alpha} \\ + 2 |R_{\mu\nu} | \cos \left[\left(\lambda_\mu - \lambda_\nu \right)\tilde{t}  + \varphi_{\mu\nu}\right] \, ,
\end{multline}
where $R_{\mu\nu} = |R_{\mu\nu}| e^{i\varphi_{\mu\nu} }   $. The initial state has no right-handed neutrinos, and $\sum_{\vec{p}} \rho_{++}(\vec{p},\tilde{t})$ cannot be negative. These conditions impose\footnote{There is no right-handed neutrino initially, such that $\sum_{\vec{p}} \rho_{++} \approx 2|R_{\mu\nu}| \left\{ \cos \left[ \left( \lambda_\mu -\lambda_\nu \right)\tilde{t} + \varphi_{\mu\nu} \right]  - \cos\varphi_{\mu\nu}  \right\}$. This is nonnegative for all $\tilde{t}$ only if $\cos\varphi_{\mu\nu}=-1$.} $\varphi_{\mu\nu}=\pi$, so we can write
\begin{equation}
\label{eq:large_oscillations}
    \sum_{\vec{p}} \rho_{++}(\vec{p},\tilde{t}) \approx 4 |R_{\mu\nu}| \sin^2 \left( \frac{\lambda_\mu - \lambda_\nu}{2} \tilde{t}\right) \, . 
\end{equation}
In the following, we will see two examples where the evolution can be described by Eq.~\eqref{eq:large_oscillations}, with two different features. First, in Figs.~\ref{fig:compare_frequencies}--\ref{fig:0p5_grid}, the amplitude $|R_{\mu\nu}|$ is independent of $\tilde{m}$, while the frequency $\propto \lambda_\mu - \lambda_\nu$ scales linearly with $\tilde{m}$. In the examples of Fig.~\ref{fig:amplitude_m2_compare_frequencies}, the amplitude is proportional to $\tilde{m}^2$, while the frequency is mass-independent.

To be more explicit, we can expand $|R_{\mu\nu}|$ in non-negative powers of $\tilde{m}$. For every $\alpha=1,\dots,k$, one has
\begin{multline}
        \left( \bar{H}_0 + \tilde{m} \bar{H}' \right) \left(\ket{\lambda^{(0)}_\alpha} + \tilde{m} \ket{\lambda^{(1)}_\alpha} + \ldots \right) \\
        = \left( \lambda^{(0)}_\alpha + \tilde{m}  \lambda^{(1)}_\alpha + \ldots \right) \left(\ket{\lambda^{(0)}_\alpha} + \tilde{m} \ket{\lambda^{(1)}_\alpha} + \ldots \right) \, ,
\end{multline}
with $\bar{H}_0 = \tilde{H}_0 + \tilde{H}_{LL\to LL} $ and $ \tilde{m}\bar{H}'= \tilde{H}_{LL\to LR} + \tilde{H}_{LR\to LL}$. At lowest order $\tilde{m}^0$, we can write
\begin{equation}
    \bar{H}_0 \ket{\lambda^{(0)}_\alpha}  = \lambda^{(0)}_\alpha \ket{\lambda^{(0)}_\alpha}   \quad , \quad   \ket{\lambda^{(0)}_\alpha} = \sum_{\beta=1}^{k} \braket{v_\beta|\lambda^{(0)}_\alpha} \ket{v_\beta}.
\end{equation}
The basis vectors $\ket{v_1},\dots,\ket{v_k}$ have the same kinetic energy, so the part of $\lambda^{(0)}_\alpha$ given by $\tilde{H}_0$ is the same for all $\alpha=1,\dots,k$. Therefore, if two eigenvalues of the full problem have a difference $\lambda_{\mu} - \lambda_{\nu}$ of order $\tilde{m}^0$, its value comes from $\tilde{H}_{LL\to LL}$. We would thus expect, in that case, oscillations with a frequency of order unity in units of $\tilde{t}$. Nevertheless, as we will see in the following, geometrical effects can lead to a frequency about a thousand times smaller. Finally, we note that $|R_{\mu\nu}|$ and $\lambda_\mu - \lambda_\nu$ cannot both be of order $\tilde{m}^0$ simultaneously, as this would mean that massless neutrinos can change helicity.

\subsubsection{Full equipartition case}

\begin{figure}[!ht]
    \centering
    \includegraphics[width=\columnwidth]{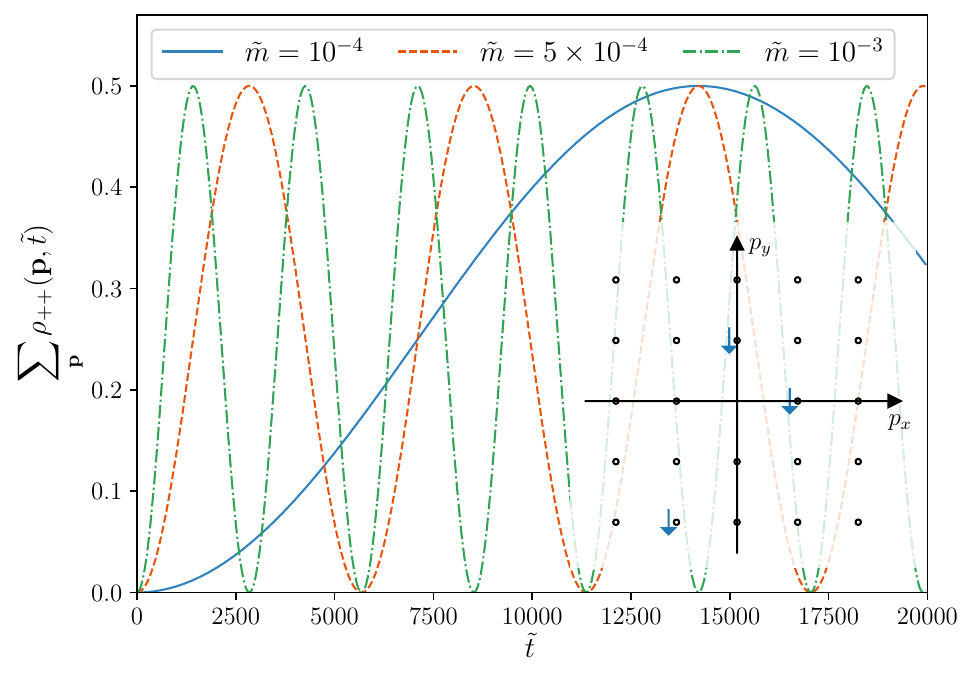}
    \caption{Evolution of the total occupation number of right-handed states for various values of the mass parameter, for the initial configuration shown in the bottom right corner. All these results correspond to full many-body calculations. The frequency of the oscillations scales linearly with $\tilde{m}$, while the maximum amplitude is, for this specific case, always $1/2$.}
    \label{fig:compare_frequencies}
\end{figure}

\begin{figure}[!ht]
    \centering
    \includegraphics[width=\columnwidth]{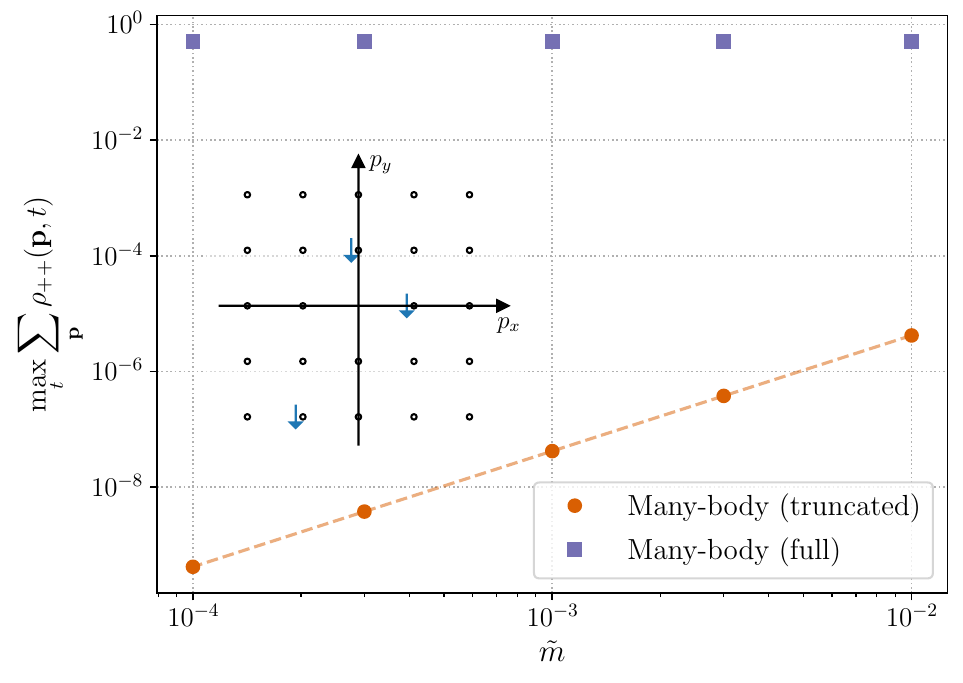}
    \caption{Maximum total right-handed occupation number (for $\tilde{t} \leq 2 \times 10^4$), for the configuration shown in Fig.~\ref{fig:compare_frequencies}. The truncated many-body (and mean-field, not shown since identical) right-handed populations follow the $\tilde{m}^2$ dependence (dashed line), when the full many-body calculation shows a periodic equipartition, as evidenced in Fig.~\ref{fig:compare_frequencies}.}
    \label{fig:loglog_0p5}
\end{figure}

We first identify an ``extreme'' case where the amplitude $|R_{\mu \nu}|$ appearing in Eq.~\eqref{eq:large_oscillations} is of order $\tilde{m}^0$. The total occupation number of right-handed states for the full many-body calculations is shown in Fig.~\ref{fig:compare_frequencies}, where we compare this quantity for different values of the mass parameter. The amplitude of the large oscillations is independent of $\tilde{m}$ and is equal to $0.5$, while the frequency is linear in $\tilde{m}$ (with for instance a doubling of the frequency between the orange and green lines). For the truncated many-body (and mean-field) calculations, there are no such large oscillations and we recover the usual behavior, as shown in Fig.~\ref{fig:loglog_0p5}.

We note that if we looked at the full many-body evolution for small times, we would still interpret the helicity conversion as scaling as $\tilde{m}^2$, as can be seen expanding Eq.~\eqref{eq:large_oscillations} at small times\footnote{We then recover the general statement of Appendix~\ref{app:m2_dependence}, which predicts a right-handed population $\propto \tilde{m}^2$, since the derivation only applies \emph{a priori} for $\tilde{m} \tilde{t} \ll 1$.} (with the difference of eigenvalues being of order $\tilde{m}$, as evidenced by the frequency dependence in Fig.~\ref{fig:compare_frequencies}). For a realistic value $\tilde{m} \lesssim 10^{-7}$, the time to reach maximum conversion would be $\tilde{t} \gtrsim 10^{7} \simeq 200 \, \mathrm{ms}$ (since, in our problem, the timescale is $t_0 \simeq 20 \, \mathrm{ns}$). This large duration needed for macroscopic spin conversion remains close to the dynamical timescales of a supernova environment, suggesting that such phenomena are not totally impossible. In addition, other spin conversion drivers (transverse matter currents, magnetic field) could further enhance the possibility of left-to-right helicity conversion.

\begin{figure}[!ht]
    \centering
    \includegraphics[width=\columnwidth]{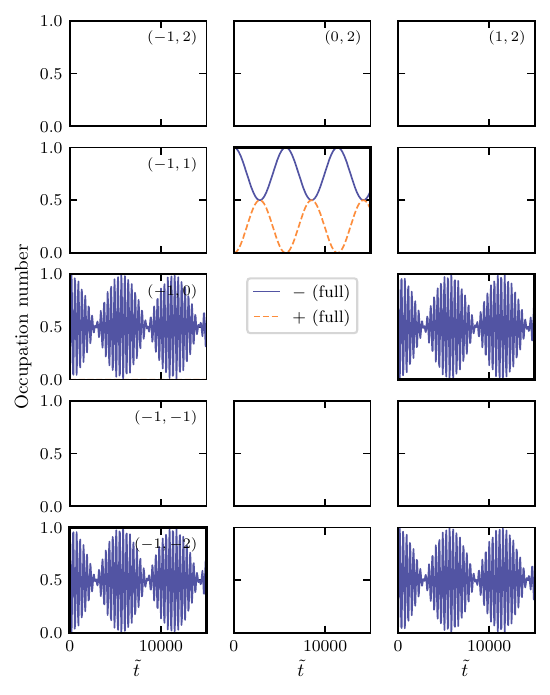}
    \caption{Same as Fig.~\ref{fig:varying_angle_grid2}, but for the peculiar configuration discussed in Figs.~\ref{fig:compare_frequencies} and \ref{fig:loglog_0p5}. The results are shown for $\tilde{m}=5 \times 10^{-4}$. We only display the panels where order unity occupation numbers are reached; there are also $\propto \tilde{m}^2$ left-handed populations in the bins $(\pm 2,-1)$.}
    \label{fig:0p5_grid}
\end{figure}

The particularly large helicity conversion observed here is a feature of one specific momentum bin, namely, $\vec{n}=(0,1)$. This is illustrated in Fig.~\ref{fig:0p5_grid}, where we use the same plotting convention as in Figs.~\ref{fig:varying_angle_grid1} and \ref{fig:varying_angle_grid2} in order to easily visualize the occupation numbers for each momentum. The left-handed populations of the initially occupied bins $(1,0)$ and $(-1,-2)$ get quickly shared with the bins $(-1,0)$ and $(2,-1)$, with an amplitude modulated by the low-frequency behavior in the bin $(0,1)$. We have found that, because of the symmetries of this configuration, the initial state is projected on a specific combination of eigenvectors of the full Hamiltonian, for which the occupation numbers in one specific momentum bin undergo large oscillations. This conclusion appears to be general: in all large-conversion configurations we have encountered, we have observed the same concentration of this large conversion to one momentum bin.

\subsubsection{Mass-independent frequency case}

\begin{figure}[!ht]
    \centering
    \includegraphics[width=\columnwidth]{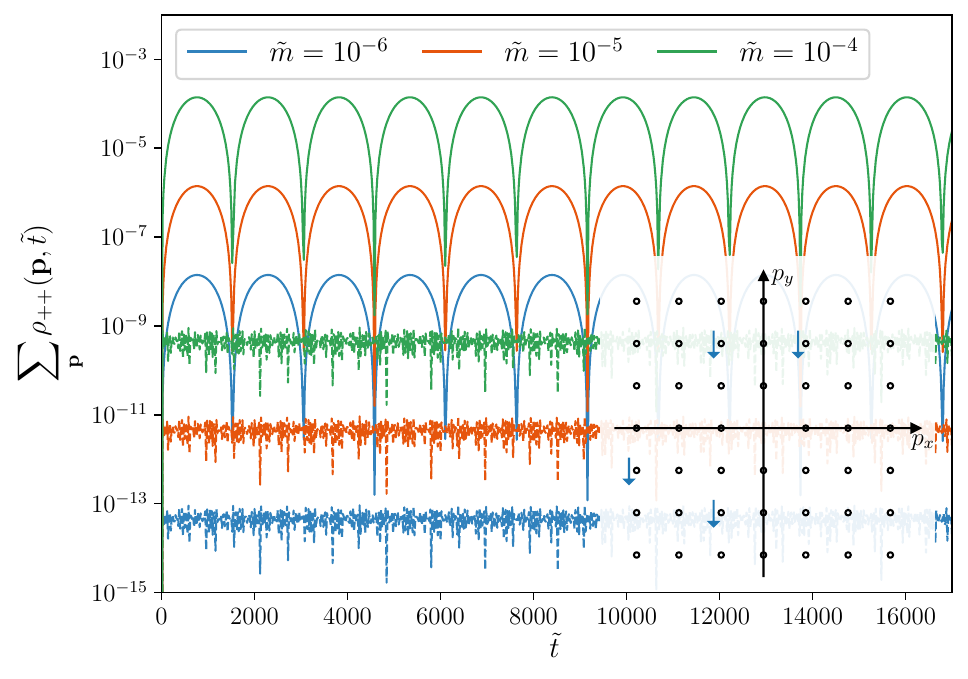}
    \includegraphics[width=\columnwidth]{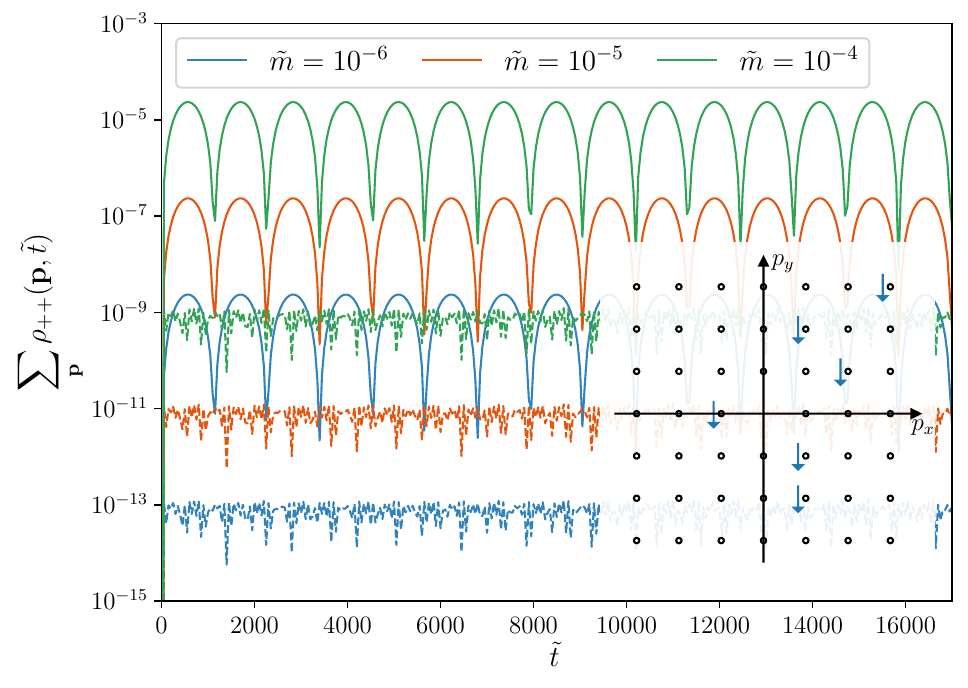}
    \caption{Evolution of the total occupation number of right-handed states for various values of the mass parameter for the configurations shown in the bottom right corners. Solid (dashed) lines correspond to full (truncated) many-body calculations.}
    \label{fig:amplitude_m2_compare_frequencies}
\end{figure}

In Fig.~\ref{fig:amplitude_m2_compare_frequencies}, we show two cases, with respectively 4 and 6 neutrinos, where it is now the frequency of oscillations that is mass-independent, but the amplitude of conversion scales as $\tilde{m}^2$. To highlight these two points, we use a logarithmic scale.

Both the full and truncated Hamiltonian many-body calculations show the $\tilde{m}^2$ dependence of $\sum_{\vec{p}}\rho_{++}(\vec{p},\tilde{t})$, but the full Hamiltonian increases left-to-right conversion by 5 (resp.~slightly more than 4) orders of magnitude for the top (resp. bottom) configuration. The low-frequency oscillations of the amplitude, which are associated with these large spin-flips, as before, are only a feature of the full Hamiltonian calculation (they correspond to the \lq\lq humps\rq\rq\ seen in Fig.~\ref{fig:amplitude_m2_compare_frequencies}). These results emphasize the diversity of situations for which the generic enhancement shown in Sec.~\ref{subsec:general_results} can be augmented by several orders of magnitude for specific configurations.

\section{Summary and Discussion}
\label{sec:summary}

In this work, we leverage recent developments on the study of collective neutrino flavor oscillations to tackle the question of neutrino \emph{helicity} conversion from a quantum many-body perspective. Several mean-field treatments in the literature have shown that such ``spin-flip'' behaviors should generally be negligible (e.g.,~\cite{Vlasenko:2014bva,Purcell:2024bim}), as expected given the standard $(\text{mass}/\text{energy})^2$ suppression of helicity conversion. Nevertheless, significant differences with mean-field results that stem from many-body entanglement have been found in various setups for neutrino flavor conversion~\cite{Patwardhan:2022mxg}. This intriguing finding motivates the present study of helicity conversion.

We use a toy model consisting of a limited two-dimensional grid of modes in momentum space, with only one generation of Dirac neutrinos of mass $m$. This allows us to ignore flavor oscillations and the presence of antineutrinos, such that the conversion we are interested in is between $\nu_L$ and $\nu_R$ states.

We compare different frameworks for neutrino evolution: the mean-field approximation and the direct solution of the $N$-body Schrödinger equation, the latter being carried out with two different Hamiltonians. First, following longstanding practices in the flavor literature, we consider a ``truncated'' Hamiltonian where only the forward and exchange terms (which are the only ones contributing at the mean-field level) are retained; we then also consider the ``full'' Hamiltonian where all momentum exchange possibilities are included (following the work of Ref.~\cite{Cirigliano:2024pnm} for flavor conversion). All expressions are derived, at leading order in the nonzero mass contribution, in Sec.~\ref{sec:evolution_equations}. For realistic environments, the kinetic energy of neutrinos is orders of magnitude larger than the strength of the two-body interaction term, which explains why kinetic energy is dynamically conserved, allowing for a further simplification of the Hamiltonian. This argument was first laid out in~\cite{Cirigliano:2024pnm}, and we revisit it using the time-dependent perturbation theory in Sec.~\ref{subsec:setup}.

For various initial momentum configurations of several left-handed neutrinos, we compare the total occupation number of right-handed neutrino states as a function of time for the different calculation frameworks. There is a systematic enhancement of left-to-right helicity conversion when many-body effects facilitated by the full Hamiltonian are included. This is the main result of this study. We attribute this effect to the additional one-particle states that are made available by the momentum-exchanging part of the full Hamiltonian. These extra states allow for an increased spin-flip probability.

Contrary to the case of neutrino flavor conversion, there are no differences between the results of mean-field and truncated Hamiltonian many-body spin-flip calculations. This is a consequence of the smallness of the parameter $m/p_0$, with $p_0 = \mathcal{O}(\mathrm{MeV})$, the typical neutrino energy (see Fig.~\ref{fig:loglog} and associated discussion). Our mean-field results are consistent with the need for a transverse potential to seed helicity conversion, as evidenced by the exploratory calculations described in Sec.~\ref{subsec:transverse_potential}. 

The helicity conversion enhancement associated with the full Hamiltonian is generally several orders of magnitude: one order of magnitude in Fig.~\ref{fig:loglog}; more than two in Fig.~\ref{fig:example_twobeams}; even four and five in the examples of Fig.~\ref{fig:amplitude_m2_compare_frequencies}. These latter configurations correspond to cases where the total right-handed occupation number displays large oscillations, as discussed in Sec.~\ref{subsec:large_conversion}. We have even identified a particularly symmetric configuration which leads to periodic equipartition of the left- and right-handed populations in one momentum bin (see Figs.~\ref{fig:compare_frequencies}--\ref{fig:0p5_grid}), noting however that the frequency of the large oscillations is then proportional to the ratio $m/p_0$.

Although the many simplifications of our toy model may lead to the particularly large spin-flip probability discussed in Sec.~\ref{subsec:large_conversion}, the fact that mean-field (and truncated many-body) calculations significantly underestimate this probability compared to full many-body calculations is a generic conclusion of our work. Several steps would need to be taken to assess the specific consequences of this mechanism in a supernova environment, including, for instance, a larger number of neutrinos, 3 mass generations, and antineutrinos. If the spin-flip enhancement carries over to the case of Majorana neutrinos, this would lead to interesting observable signatures, as the process would effectively correspond to a neutrino $\to$ antineutrino conversion. 

That might be visible, for instance, in the neutronization burst of a supernova. That burst would have prodigious luminosity for roughly $\sim 10\,{\rm ms}$. We also would expect the $\nu_e$ flux in the burst to exceed the fluxes of other neutrino and antineutrino types by roughly an order of magnitude. Flavor transformation in the envelope in the normal mass hierarchy will convert some of the $\nu_e$s to other flavors. However, absent large scale spin-flip of Majorana $\nu_e$s, we would not expect significant numbers of $\bar\nu_e$s to appear. That is intriguing. Should an anomalously large $\bar\nu_e$ flux from a Galactic core-collapse event be observed in a terrestrial detector, there would be speculation about spin-flip. This might arise from medium effects, including the many-body enhancement evidenced in this work, or from a large neutrino magnetic moment as mentioned in the introduction.  

However, our toy model, while suggestive of significant many-body enhancement to spin-flip, has obvious and significant limitations as outlined above.  Nevertheless, note that only neutrinos are present in our model, while matter effects or interactions with external magnetic fields might contribute further. These other spin-flip causes were usually considered as the dominant sources of helicity conversion (e.g.,~\cite{deGouvea:2012hg,Chatelain:2016xva,Purcell:2024bim}). We conjecture that many-body neutrino effects, in conjunction with these standard mechanisms, might further increase the spin-flip probability.

Finally, the particular setup we considered, with neutrinos in well-defined momentum states, has been criticized in the context of flavor conversion since it would lead to too many interactions (and therefore quantum correlations and entanglement) compared to the standard quantum kinetic approach~\cite{Shalgar:2023ooi,Johns:2023ewj}, which builds upon a separation of scales not satisfied here. The same arguments apply in the case of helicity conversion. Extensions of this work towards more realistic wave packet~\cite{Cervia:2025pfg} descriptions, or intermediate approaches between the mean-field and many-body limits (e.g.,~\cite{Laraib:2025uza,Laraib:2025ziz,Kost:2024esc,Kost:2025vyt}), could be pursued in the future. Given the stakes for compact object physics, and given the intriguing many-body enhancement found here, we conclude that further exploration of neutrino spin-flip is warranted.

\begin{acknowledgments}
We thank V. Cirigliano, S. Sen and Y. Yamauchi for discussions on the code they developed, described in Ref.~\cite{Cirigliano:2024pnm}. We also thank Baha Balantekin for numerous discussions on many-body neutrino physics. Y.X. thanks Yueqi Zhao for useful discussions. J.F. acknowledges support from the Severo Ochoa Excellence Grant CEX2023-001292-S funded by MICIU/AEI/10.13039/501100011033. This work was supported in part by National Science Foundation (NSF) Grants  No.~PHY-2209578 and No.~PHY-2515110 at UCSD and the {\it Network for Neutrinos, Nuclear Astrophysics, and Symmetries} (N3AS) NSF Physics Frontier Center, NSF Grant No.~PHY-2020275, the Heising-Simons Foundation (2017-228), and the UC San Diego Academic Senate. L.G. acknowledges support from the Dutch Research Council (NWO) under project number VI.Veni.222.318 and from Charles University through the project number
PRIMUS/24/SCI/013.
\end{acknowledgments}

\appendix

\section{Quantized Dirac field and weak interactions}
\label{app:QFT}

\subsection{Neutrino Hamiltonian}
\label{app:interaction_picture}

We start by rigorously deriving the expression of the neutrino Hamiltonian. The goal is to provide a self-contained bridge between the four-fermion Lagrangian, the interaction-picture field expansion in a finite volume, and the operator structures that appear in the spin/helicity-resolved Hamiltonians used in our numerical many-body and mean-field calculations. As in the main text, we work with a single Dirac neutrino species of mass $m$ interacting via the Standard-Model neutral-current (V--A) four-fermion operator in the low-energy limit. We use the metric $\eta_{\mu\nu}=\mathrm{diag}(+,-,-,-)$.

The effective Lagrangian density is 
\begin{equation}
    \mathcal{L} = \bar{\psi}\left(i\gamma^\sigma\partial_\sigma-m\right)\psi - \frac{G_F}{\sqrt{2}} \left(\bar{\psi}\gamma^\sigma P_L\psi \right)\left(\bar{\psi}\gamma_\sigma P_L\psi \right),
\end{equation}
where $P_L = (1-\gamma^5)/2$ is the
left-handed chiral projector. $\psi$ is the Dirac spinor for the neutrino of mass $m$. The quantum Heisenberg-picture Hamiltonian $H^H(t)$ can be obtained by using the canonical quantization procedure (see e.g.,~\cite{Greiner:1996zu,Neto:2021hhl}). 

The canonical Hamiltonian at time $t$ is obtained from
\begin{equation}
    H(t) = \int \dd^3x \left(\pi_\psi \dot{\psi} + \pi_{\psi^\dagger} \dot{\psi}^\dagger -\mathcal{L}\right),
\end{equation}
with conjugate momenta $\pi_\psi = \partial\mathcal{L}/\partial \dot{\psi}$ and $\pi_{\psi^\dagger} = \partial\mathcal{L}/\partial \dot{\psi}^\dagger$. Canonical quantization promotes the fields to operators and imposes equal-time anticommutation relations. In the Heisenberg picture we introduce the field operators $\nu^H(x)$ and ${\nu^H}^\dagger(x)$ which satisfy
\begin{equation}
\begin{aligned}
    \left\{ \nu^H_a(\vec{x},t), {\nu^H}^\dagger_b(\vec{x}',t)\right\} &= \delta_{ab} \delta^{(3)}(\vec{x}-\vec{x}') \, , \label{eq:app_equal_time} \\
    \left\{ \nu^H_a(\vec{x},t), \nu^H_b(\vec{x}',t)\right\} &= 0 \, , \\
    \left\{ {\nu^H}^\dagger_a(\vec{x},t), {\nu^H}^\dagger_b(\vec{x}',t)\right\} &= 0 \, .
\end{aligned}
\end{equation}
Here, $a,b\in\{1,2,3,4\}$ label the Dirac spinor components. Replacing the Dirac spinors $\psi(x)$ and $\psi^\dagger(x)$ by the field operators $\nu^H(x)$ and ${\nu^H}^\dagger(x)$ in the canonical Hamiltonian gives the Heisenberg-picture Hamiltonian $H^H(t)$.\footnote{In standard cases such as the closed neutrino system we study here, the Hamiltonian is not explicitly time-dependent and $H^H(t) = H^H(0)$ at all times (as can be seen applying the Heisenberg equation). We keep the discussion here more general, so that it can be adapted to other cases as well.}

It is convenient to split
\begin{equation}
    H^H(t) = H^H_0(t) + H^H_{\mathrm{int}}(t),
\end{equation}
where $H^H_0$ is the free Dirac Hamiltonian and $H^H_{\mathrm{int}}$ contains the four-fermion interaction. In what follows, the Schr\"odinger-picture Hamiltonian is $H \equiv H^H(0)$.

We define the interaction-picture field by evolving the Heisenberg field with the free Hamiltonian,
\begin{equation}
    \nu^I(x) = e^{iH^H_0(0) t}\nu^H(\vec{x},0)e^{-iH^H_0(0)t}.
\end{equation}
$H^H_0(t)$ is the vacuum part of $H^H(t)$.
By construction, $\nu^I(x)$ satisfies the free Dirac equation,
\begin{equation}
\left(i\gamma^\sigma\partial_\sigma-m\right)\nu^I(x)=0,
\end{equation}
and we use $E_{\vec p}=\sqrt{p^2+m^2}$ with $p\equiv|\vec p|$.

The interaction-picture field can be expanded as
\begin{multline}
    \nu^I(x) = \sum_{\vec{p},h}  \frac{1}{\sqrt{2VE_{\vec{p}}}}\bigg[a_h(\vec{p}) u_h(\vec{p}) e^{-ipx} \\ 
    +
    b^\dagger_h(\vec{p}) v_h(\vec{p}) e^{ipx} \bigg],
    \label{eq:app_field_expansion}
\end{multline}
where $p^\mu=(E_{\vec p},\vec p)$, $x^\mu=(t,\vec x)$. 
The label $h=\pm$ denotes helicity, i.e. eigenvalues of $\bm{\Sigma}\cdot\hat p$ where $\bm{\Sigma}=\gamma^5\gamma^0\bm{\gamma}$ (equivalently of $\bm{\sigma}\cdot\hat p$ on the two-component spinors used below). The operators $a_h(\vec p)$ and $b_h(\vec p)$ annihilate a neutrino and an antineutrino with momentum $\vec p$ and helicity $h$, respectively.

The non-vanishing anticommutation relations of the creation and annihilation operators are
\begin{equation}
    \left\{ a_h(\vec{p}), a^\dagger_{h'}(\vpp)\right\} = \left\{ b_h(\vec{p}), b^\dagger_{h'}(\vpp)\right\} =\delta_{hh'} \delta_{\vec{p}\vpp}.
\end{equation}
In the discrete box normalization,
\begin{equation}
    \delta_{\vec{p}\vec{p}\,'}
    =
    \delta_{n_x n'_x}\,\delta_{n_y n'_y}\,\delta_{n_z n'_z},
    \qquad
    \vec p=\frac{2\pi}{L}\vec n,\ \ \vec p\,'=\frac{2\pi}{L}\vec n'.
\end{equation}
Together with the spinor completeness/orthonormality relations in Appendix~\ref{app:spinors}, Eq.~\eqref{eq:app_field_expansion} reproduces the equal-time field anticommutators in Eq.~\eqref{eq:app_equal_time}.

Finally, the Schr\"odinger-picture field operator at $t=0$ is given as
\begin{equation}
    \nu(\vec x)\equiv \nu^H(\vec x,0)=\nu^I(\vec x,0) \, ,
\end{equation}
matching the expansion~\eqref{eq:Fourier_expansion} in the main text. The interaction Hamiltonian in the Schr\"odinger picture follows directly from the four-fermion term in $\mathcal{L}$,
\begin{equation}
    H_{\mathrm{int}}
    =
    \frac{G_F}{\sqrt{2}}
    \int \dd^3 \vec{x}\,
    \left[\bar{\nu}(\vec x)\gamma^\mu P_L \nu(\vec x)\right]
    \left[\bar{\nu}(\vec x)\gamma_\mu P_L \nu(\vec x)\right] \, .
    \label{eq:app_hint}
\end{equation}
Inserting the mode expansion of $\nu(\vec x)$ yields a quartic operator expressed in terms of $a,a^\dagger,b,b^\dagger$ and spinor bilinears of the form $\bar{u}_{h_1}(\vec p_1)\gamma^\mu P_L u_{h_2}(\vec p_2)$, $\bar{u}_{h_1}(\vec p_1)\gamma^\mu P_L v_{h_2}(\vec p_2)$, etc. The spatial integral produces momentum conservation,
\begin{equation}
    \int \dd^3 \vec{x}\,e^{i(\vec p_1+\vec p_2-\vec p_3-\vec p_4)\cdot \vec x}
    = V\,\delta_{\vec p_1+\vec p_2,\ \vec p_3+\vec p_4}.
    \label{eq:app_momentum_delta}
\end{equation}
Equation~\eqref{eq:app_momentum_delta} implies that the interaction couples momentum modes only in combinations that preserve total momentum. In the main text we specialize to neutrinos (discarding antineutrino operators) as a controlled simplification and retain the relevant terms at the desired order in $m/E$ for helicity conversion. The explicit helicity kernels appearing in the reduced Hamiltonians are built from the spinor bilinears above and are evaluated with the conventions given below in Appendix~\ref{app:spinors}.

\subsection{Spinor conventions}
\label{app:spinors}

\subsubsection{Gamma matrices and two-component helicity spinors}
\label{app:spinors:gammas}

We use the chiral/Weyl representation of the Dirac matrices,
\begin{equation}
\begin{aligned}
    &\gamma^0 =
    \begin{pmatrix}
        0 & \mathbb{I}_2 \\
        \mathbb{I}_2 & 0
    \end{pmatrix},
    \qquad
    \bm{\gamma} =
    \begin{pmatrix}
        0 & \bm{\sigma} \\
        -\bm{\sigma} & 0
    \end{pmatrix},
    \\
    &\gamma^5 \equiv i\gamma^0\gamma^1\gamma^2\gamma^3
    =
    \begin{pmatrix}
        -\mathbb{I}_2 & 0 \\
        0 & \mathbb{I}_2
    \end{pmatrix},
\end{aligned}
\end{equation}
where $\bm{\sigma} = (\sigma_1, \sigma_2, \sigma_3)$ is the vector of Pauli matrices, and $\mathbb{I}_2$ is the $2\times2$ identity.

For any momentum direction $\hat p\equiv \vec p/p$, the two-component helicity eigenspinors $\xi_h(\hat p)$ are the eigenvectors of $\bm{\sigma}\cdot \hat p$,
\begin{equation}
    (\bm{\sigma}\cdot \hat p)\,\xi_h(\hat p)=h\,\xi_h(\hat p),
    \qquad h=\pm 1.
\end{equation}
Writing $\hat p$ in spherical angles $(\theta_{\vec p},\phi_{\vec p})$, one convenient choice is (this is the same convention as~\cite{Cirigliano:2014aoa})
\begin{equation}
    \xi_+(\hat{p}) =
    \begin{pmatrix}
        \cos \frac{\theta_{\vec{p}}}{2}   \\
        e^{i\phi_{\vec{p}}}\sin \frac{\theta_{\vec{p}}}{2}
    \end{pmatrix},
    \qquad
    \xi_-(\hat{p}) =
    \begin{pmatrix}
        -e^{-i\phi_{\vec{p}}} \sin \frac{\theta_{\vec{p}}}{2} \\
        \cos \frac{\theta_{\vec{p}}}{2}
    \end{pmatrix}.
    \label{eq:app_xi_pm}
\end{equation}
These satisfy $\xi_h^\dagger(\hat p)\xi_{h'}(\hat p)=\delta_{hh'}$ and the completeness relation $\sum_h \xi_h(\hat p)\xi_h^\dagger(\hat p)=\mathbb{I}_2$.

\subsubsection{Dirac spinors and normalization}
\label{app:spinors:dirac_spinors}

Introducing
\begin{equation}
    r(p)\equiv \frac{m}{E_{\vec p}+p},
    \qquad E_{\vec p}=\sqrt{p^2+m^2} \, ,
\end{equation}
such that $r(p)\simeq m/(2p)$ in the ultrarelativistic limit, and using the conventions~\eqref{eq:app_xi_pm}, the positive-energy (neutrino) helicity spinors are
\begin{equation}
\begin{aligned}
    u_+(\vec{p}) &= \sqrt{E_{\vec{p}}+p}\,
    \begin{pmatrix}
        r(p)\,\xi_+(\hat{p}) \\
        \xi_+(\hat{p})
    \end{pmatrix},
    \\
    u_-(\vec{p}) &= \sqrt{E_{\vec{p}}+p}\,
    \begin{pmatrix}
        \xi_-(\hat{p}) \\
        r(p)\,\xi_-(\hat{p})
    \end{pmatrix}.
    \label{eq:app_u_spinors}
\end{aligned}
\end{equation}
The negative-energy (antineutrino) helicity spinors are taken as\footnote{Although we do not consider antineutrinos in this work, we include these definitions for completeness.}
\begin{equation}
\begin{aligned}
    v_+(\vec{p}) &= \sqrt{E_{\vec{p}}+p}\,
    \begin{pmatrix}
        \xi_-(\hat{p}) \\
        -r(p)\,\xi_-(\hat{p})
    \end{pmatrix},
    \\
    v_-(\vec{p}) &= \sqrt{E_{\vec{p}}+p}\,
    \begin{pmatrix}
        -r(p)\,\xi_+(\hat{p}) \\
        \xi_+(\hat{p})
    \end{pmatrix}.
    \label{eq:app_v_spinors}
\end{aligned}
\end{equation}
With these conventions the usual orthonormality relations hold,
\begin{equation}
\begin{aligned}
    u_h^\dagger(\vec p)\,u_{h'}(\vec p) &= 2E_{\vec p}\,\delta_{hh'},
    \\
    v_h^\dagger(\vec p)\,v_{h'}(\vec p) &= 2E_{\vec p}\,\delta_{hh'},
    \\
    u_h^\dagger(\vec p)\,v_{h'}(\vec p) &= 0,
    \\
    \bar u_h(\vec p)\,u_{h'}(\vec p) &= 2m\,\delta_{hh'},
    \\
    \bar v_h(\vec p)\,v_{h'}(\vec p) &= -2m\,\delta_{hh'}, & &
\end{aligned}
\end{equation}
along with the completeness relations
\begin{equation}
\begin{aligned}
    \sum_h u_h(\vec p)\bar u_h(\vec p)&=\slashed{p} + m \, ,
    \\
    \sum_h v_h(\vec p)\bar v_h(\vec p)&=\slashed{p} - m \, .
\end{aligned}
\end{equation}
These identities ensure that the field expansion~\eqref{eq:app_field_expansion} is consistent with the equal-time anticommutation relations.

\section{Algorithm details}
\label{app:numerical_details}

In this Appendix, we explain in detail our numerical construction of the many-body neutrino Hamiltonian, which follows the strategy of~\cite{Cirigliano:2024pnm}, adapted for helicity conversion with only one flavor. 

\begin{figure}[!ht]
    \centering
    \includegraphics[trim={0 0.5cm 0 0.3cm},clip,width=\columnwidth]{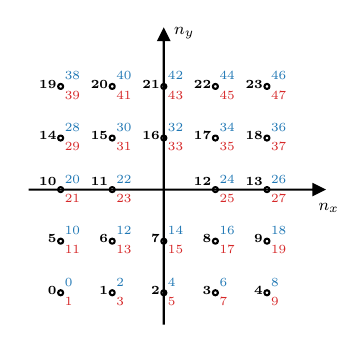}
    \caption{Example of momentum grid with a maximum quantization index $n_\mathrm{max} = 2$. The numbering of the momentum states $l = 0, \dots, 4 n_\mathrm{max} (n_\mathrm{max} + 1)-1$ is shown in bold black, while the even (odd) numbering of the one-body helicity states is shown in blue (red) for left-handed (right-handed) states.}
    \label{fig:numbered_lattice}
\end{figure}
    
To make things explicit, we consider the momentum grid represented in Fig.~\ref{fig:numbered_lattice} as an example—it corresponds to $n_\text{max}=2$ in Eq.~\eqref{eq:2D_grid}. There are $24$ momentum modes (we exclude the zero-momentum mode) labeled in an increasing order of $n_x$ and $n_y$, for instance
\begin{equation}
\begin{aligned}
    \vec{n}_0 &= (-2,-2), & \vec{n}_1 &= (-1, -2), & \vec{n}_4 &= (2, -2), \\
    \vec{n}_5 &= (-2,-1), & \vec{n}_{13} &= (2,0), & \vec{n}_{23} &= (2,2).
\end{aligned}
\end{equation}
Since each momentum mode can be associated with two helicities, there are $48$ one-particle states. For any momentum mode $\vec{n}_l$ of $l=0,\dots,23$, the left-handed state is represented by the integer $2l$ and the right-handed one by $2l+1$. Some one-particle states are thus
\begin{equation}
\begin{aligned}
    0 &\leftrightarrow \begin{pmatrix}
        -2 \\ -2
    \end{pmatrix}_-, & 
    1 &\leftrightarrow \begin{pmatrix}
        -2 \\ -2
    \end{pmatrix}_+, & 
    8 &\leftrightarrow \begin{pmatrix}
        2 \\ -2
    \end{pmatrix}_-   \\
    10 &\leftrightarrow \begin{pmatrix}
        -2 \\ -1
    \end{pmatrix}_-, & 
    27 &\leftrightarrow \begin{pmatrix}
        2 \\ 0
    \end{pmatrix}_+, &
    47 &\leftrightarrow \begin{pmatrix}
        2 \\ 2
    \end{pmatrix}_+.
\end{aligned}
\end{equation}
Therefore, the set of discrete creation operators is
\begin{equation}
    a^\dagger_{2l} = a^\dagger_-(p_0 \vec{n}_l) \quad ,\quad a^\dagger_{2l+1} = a^\dagger_+(p_0\vec{n}_l)
\end{equation}
for $l=0,\dots, 23$. To avoid the ambiguity of ordering the creation operators $a^\dagger_h(\vec{p})$ for the eigenstates $\ket{n}$ of $H_0$, we take the following convention;
\begin{equation}
\label{eq:mode_ordered}
    \ket{i_1,\dots, i_N} = a^\dagger_{i_1}\dots a^\dagger_{i_N}\ket{0} \, ,
\end{equation}
for $0 \leq i_1 < \dots < i_N \leq 47$. 

The dimensionless Hamiltonian~\eqref{eq:Hamiltonian_dimensionless} is 
\begin{align}
    \tilde{H}_0 &= \frac{p_0}{\mu} \sum_{i=0}^{47} |\vec{n}_{\bar{\imath}}| a^\dagger_i a_i \, , \\
    \tilde{H}_{\nu\nu}^{(1)} &= -4 \sum_{i',f'=1}^{47} \sum_{i<i', f<f'} g( f, i ,f' , i') \, a^\dagger_f a^\dagger_{f'} a_i a_{i'} \nonumber \\
    &\qquad \times \delta_{\vec{n}_{\bar{\imath}} + \vec{n}_{\bar{\imath}'}, \vec{n}_{\bar{f}}+\vec{n}_{\bar{f}'}} \, \delta_{|\vec{n}_{\bar{\imath}}| + |\vec{n}_{\bar{\imath}'}|, |\vec{n}_{\bar{f}}|+|\vec{n}_{\bar{f}'}|} \, ,
\end{align}
where we wrote $\bar{\imath} \equiv \lfloor i/2 \rfloor$ for conciseness. The restrictions $i<i'$ and $f<f'$ for $\tilde{H}_{\nu\nu}^{(1)}$ follow from the definition in \eqref{eq:mode_ordered} and are responsible for the prefactor $4$.

In order to diagonalize the Hamiltonian, we need to construct its matrix representation on a given basis. If the initial state has $N$ neutrinos, a complete basis for the problem consists of the states~\eqref{eq:mode_ordered}. However, not all $N$-particle states are ``accessible'' from a given $\ket{\Psi_0}$, because of energy and momentum conservation. For each initial state, we thus construct the minimal basis of $N$-particle states.

To do so, we first create a table of all possible two-body scatterings $\ket{i,i'} \to \ket{f,f'}$ which conserve energy and momentum, and for which $g(f,i,f', i') \neq 0$. For an initial state $\ket{\Psi_0}$ of the form \eqref{eq:mode_ordered}, the action of $\tilde{H}_{\nu\nu}^{(1)}\ket{\Psi_0}$ is to replace any two integers among $i_1$ to $i_N$ with a couple $(f,f')$ from the scattering table previously constructed. We thus obtain a set of vectors of the form~\eqref{eq:mode_ordered} that we call $S_1$. We then apply $\tilde{H}_{\nu\nu}^{(1)}$ to each vector in $S_1$, which induces a set $S_2$. After multiple iterations, no new vector appears in $S$ (i.e., there is an integer $m$ such that $S_m \subset \bigcup_{j < m} S_j$). The set of vectors in $\bigcup_{j < m} S_j$ are the basis states $\ket{v_1}, \dots, \ket{v_k}$. We can thus represent $\tilde{H}$ as a $k \times k$ matrix on this basis, and diagonalize it.

\section{Scaling $\propto \tilde{m}^2$ of the right-handed population}
\label{app:m2_dependence}

We can justify the fact that $\sum_{\vec{p}} \rho_{++}(\vec{p}) \propto \tilde{m}^2$, observed in the main text (e.g, Figs.~\ref{fig:loglog} and \ref{fig:example_twobeams}), by using perturbation theory. We have already used a similar argument to justify pairwise kinetic energy conservation in Sec.~\ref{subsec:kinetic_energy}, and we consider now the ``unperturbed Hamiltonian'' $\tilde{H}_0 + \tilde{H}_{LL \to LL}$ and the ``perturbation'' $\tilde{H}_{LL \to LR} + \tilde{H}_{LR \to LL} \equiv \tilde{H}_{LL \leftrightarrow LR}$, with the perturbation parameter $\tilde{m} \ll 1$. 

The unperturbed Hamiltonian conserves the number of right-handed neutrinos in every momentum,
\begin{equation}
    \left[ a^\dagger_+(\vec{k})a_+(\vec{k}) , \tilde{H}_0 + \tilde{H}_{LL \to LL} \right] = 0 \, .
\end{equation}
Since they commute, the Hamiltonian and the right-handed number operators can be simultaneously diagonalized:
\begin{equation}
\begin{aligned}
    \left( \tilde{H}_0 + \tilde{H}_{LL \to LL} \right) \ket{\lambda,n_{\vec{p}} , n_{\vec{q}} ,... } &= \lambda \ket{\lambda, n_{\vec{p}}  , n_{\vec{q}} ,...}   \\
    a^\dagger_+(\vec{k})a_+(\vec{k})  \ket{\lambda, n_{\vec{p}} , n_{\vec{q}} , ...} &= n_{\vec{k}} \ket{\lambda, n_{\vec{p}} , n_{\vec{q}} , ...} 
\end{aligned}
\end{equation}
with $n_{\vec{k}} \in \{0,1\}$ the right-handed occupation number of each momentum bin. We consider initial states of the form
\begin{equation}
\label{eq:psi0_app}
\ket{\Psi(0)}= a^\dagger_-(\vec{q}_1) \dots a^\dagger_-(\vec{q}_L) a^\dagger_+(\vec{p}_1)\dots a^\dagger_+(\vec{p}_R) \ket{0} \, .
\end{equation}
The wavefunction in \eqref{eq:Schrodinger_dimensionless} can be written
\begin{equation}
    \ket{\Psi(\tilde{t})} = \sum_{\lambda, n_{\vec{p}}, n_{\vec{q}}, \dots} c_{\lambda, n_{\vec{p}} , n_{\vec{q}} ,\dots}(\tilde{t}) e^{-i\lambda \tilde{t}} \ket{\lambda, n_{\vec{p}} , n_{\vec{q}} \, , \dots} \, ,
\end{equation}
and the right-handed neutrino occupation number for momentum $\vec{p}$ is 
\begin{multline}
\label{eq:rho++_rho0}
    \rho_{++}(\vec{p},\tilde{t}) - \rho_{++}(\vec{p},0) = \\
    \sum_{\lambda, n_{\vec{q}}, \dots } \left(\left| c_{\lambda, 1 , n_{\vec{q}} ,\dots } (\tilde{t}) \right|^2 - \left| c_{\lambda, 1 , n_{\vec{q}},\dots } (0) \right|^2 \right) \, .
\end{multline}
We can write at first order in the perturbation $\tilde{H}_{LL \leftrightarrow LR}$
\begin{multline}
\label{eq:d_dt_c_LR}
    i\frac{\dd c_{\lambda, n_{\vec{p}} , n_{\vec{q}} , \dots}}{\dd \tilde{t}} \approx \sum_{\lambda', n'_{\vec{p}} , n'_{\vec{q}} , \dots} c_{\lambda', n'_{\vec{p}} , n'_{\vec{q}} , \dots} (0) e^{i\left( \lambda-\lambda'\right)\tilde{t}} \\  \braket{\lambda, n_{\vec{p}} , n_{\vec{q}} , \dots| \tilde{H}_{LL \leftrightarrow LR}|\lambda', n'_{\vec{p}} , n'_{\vec{q}} , \dots} \, .
\end{multline}
Since $\lVert \tilde{H}_{LL \leftrightarrow LR} \rVert = \mathcal{O}(\tilde{m})$ [see Eqs.~\eqref{eq:H_LL_LR}--\eqref{eq:H_LR_LL}], this expansion is valid for $\tilde{m} \tilde{t} \ll 1$. We then distinguish between two cases.

For $c_{\lambda, n_{\vec{p}} , n_{\vec{q}} , \dots}(0) =0$, integration of~\eqref{eq:d_dt_c_LR} gives
\begin{equation}
    c_{\lambda, n_{\vec{p}} , n_{\vec{q}} ,\dots} (\tilde{t}) \approx \mathcal{O}(\tilde{m}) \, ,
\end{equation}
and that gives a $\propto \tilde{m}^2$ contribution to \eqref{eq:rho++_rho0}.

For nonvanishing $c_{\lambda,n_{\vec{p}} ,n_{\vec{q}} ,\dots} (0)$, only the other non-vanishing $c_{\lambda',n'_{\vec{p}},n'_{\vec{q}},\dots }(0) \neq 0$ contribute on the right side of \eqref{eq:d_dt_c_LR}. Given the initial state~\eqref{eq:psi0_app}, the coefficients which don't vanish initially, $c_{\lambda, n_{\vec{p}}, n_{\vec{q}}, \dots}(0) = \braket{\lambda, n_{\vec{p}} , n_{\vec{q}} , \dots|\Psi(0)} \neq 0$, are such that $n_{\vec{k}}= 1 $ for $\vec{k} \in \{\vec{p}_1,\dots,\vec{p}_R\}$ and 0 otherwise. The sum in \eqref{eq:d_dt_c_LR} is thus restricted to a subspace of fixed right-handed occupation numbers... but the perturbation $\tilde{H}_{LL \leftrightarrow LR}$ precisely changes those numbers by one unit. Therefore, the right-hand side of \eqref{eq:d_dt_c_LR} vanishes, and the variation of $c_{\lambda,n_{\vec{p}} ,n_{\vec{q}},\dots}$ is of second order in perturbation theory. Namely,
\begin{equation}
    c_{\lambda,n_{\vec{p}},n_{\vec{q}},\dots }(\tilde{t}) \approx c_{\lambda,n_{\vec{p}} ,n_{\vec{q}}  ,\dots}(0) + \mathcal{O}(\tilde{m}^2) \, .
\end{equation}
In that case, the contribution to $\rho_{++}(\vec{p},\tilde{t}) - \rho_{++}(\vec{p},0) $ is also $\mathcal{O}(\tilde{m}^2)$. 

\bibliography{references}

\end{document}